\documentstyle[prb,aps,floats,psfig]{revtex} 

\def\figwidth{10cm}
\long\def\levfig#1{#1}

\def\tp{\hbox{$t_{\perp}$} }         
\def\ww{\hbox{${\cal E}_{\para}$} }          
\def\td{\hbox{${\cal T}_{\perp}$} }          
\def\cp{\hbox{$c_{\perp}$} }         
\def\tpal{\hbox{$t_{eff}$} }         
\def\teff{\tpal}             
\def\R{{\cal R}}
\def\G{{\cal G }}

\def\para{\parallel }

\def\resc{{\sqrt{D'}} }     

\def\reg{\downarrow}
\def\Reg{\Downarrow}

\def\D{{\cal D} }

\def\kom{\v k}
\def\xt{\v x}  
\def\yt{\v y}  
\def\rt{\v r}  

\def\xx{{x_{\para}}}
\def\xy{{x_{\perp}}}

\def\FF{{\cal F}}

\def\PP{{\cal P}}

\def\fr{\hbox{$\FF_0$}}  
\def\afr{f}                   
\def\bfr{\hbox{$\bar{\FF_0}$}}  
\def\abfr{\bar f}                   

\def\om{\omega}
\def\iom{i \omega}
\def\kx{{k_{\para}}}
\def\ky{{k_{\perp}}}
\def\g{\sigma}

\def\v#1{{\bf #1}}

\def\vv{\v v}

\def\be{\beta}

\def\al{\alpha}
\def\th{\theta}

\def\aal{\mu}
\def\bbe{\nu}

\def\eps{\varepsilon}

\def\grad{\v{\nabla}}

\def\del{\delta}

\def\Tc{T$_c$ }

\def\D{{\cal D}}

\def\sign{\,{\rm sign}\,}

\def\lam{\lambda}

\def\fc#1{\left| #1 \right|}

\def\LO{LO }

\def\QQ{K_{\rho}}                

\def\de{\partial}

\long\def\taglia#1{#1}

\long\def\tagliasi#1{}

\newcommand{\cmag}{\gtrsim}

\def\beq{\begin{equation}}
\def\eeq{\end{equation}}
\def\beqn{\begin{eqnarray}}
\def\eeqn{\end{eqnarray}}

\def\eqref#1{ 
 (\ref{#1})}

\long\def\singlecol#1{
\twocolumn[\hsize\textwidth\columnwidth\hsize\csname @twocolumnfalse\endcsname
              #1]}

\ifpreprintsty 
\long\def\singlecol#1{#1}
\fi 

\long\def\singlecol#1{#1}

\def\eqref#1{ Eq.~(\ref{#1})}

\def\check#1{\marginpar{\rule{1mm}{10mm}} {\bf #1 }}

\long\def\taglia#1{}

\begin{document}  
\draft   

\title{ 
 Crossover to Fermi-liquid behavior 
 for weakly--coupled Luttinger liquids \\
in the anisotropic large--dimension limit
}

\author{E. Arrigoni 
}

\address{ 
Institut f\"ur Theoretische Physik,
Universit\"at W\"urzburg,
D-97074 W\"urzburg, Germany \\
e-mail: arrigoni@physik.uni-wuerzburg.de 
}

\singlecol{
 \date{\today} 
\maketitle
\begin{abstract}

We study the problem of the crossover from one- to higher-dimensional
metals by considering an array of Luttinger liquids (one-dimensional
chains) coupled by a weak interchain hopping \mbox{\tp.}  We evaluate
the exact asymptotic low--energy behavior of the self--energy in the
anisotropic infinite--dimension limit.  This limit extends the
dynamical mean field concept to the case of a chain embedded in a
self--consistent medium.  The system flows to a Fermi--liquid fixed
point for energies below the dimensional crossover temperature, and
the anomalous exponent $\al$ renormalizes to zero, in the case of
equal spin and charge velocities.  In particular, the single--particle
spectral function shows sharp quasiparticle peaks with nonvanishing
weight along the whole Fermi surface, in contrast to the lowest--order
result.  Our result is obtained by carrying out a resummation of all
diagrams of the expansion in \tp contributing to the anisotropic
$D\to\infty$ limit.  This is done by solving, in an almost completely
analytic way, an asymptotically exact recursive equation for the
renormalized vertices, within a skeleton expansion.  Our outcome shows
that perturbation expansions in \tp restricted to lowest orders are
unreliable below the crossover temperature.  The extension to finite
dimensions is discussed.  This work extends our recent Letter
[Phys. Rev. Lett. {\bf 83}, 128 (1999)], and includes all mathematical
details.

\end{abstract}
}

\section{Introduction}


According to Fermi-liquid (FL) theory~\cite{land.56,pines.nozie},
a quasiparticle is identified by a single dispersive coherent
peak  in 
  the single-particle spectral function 
describing a particle or a hole
close to the Fermi surface (FS).
This peak  becomes sharper when
approaching the FS, which reflects the fact that the lifetime of the
quasiparticle becomes infinite at the FS, while keeping its total weight $Z$
 (quasiparticle weight) finite.
On the other hand, FL theory fails generically in one
dimension, where quasiparticles  are not well defined,
and the
elementary excitations consist of collective charge and spin 
excitations with bosonic properties.
In this case,
the single-particle spectral function
 shows two dispersing peaks, corresponding to charge and spin
modes.
The splitting into 
two peaks corresponds to the decay of the quasiparticle into spin
and charge excitations~\cite{me.sc.92,voit.93,voit.rev,diagnostic.97},
i. e., the spin and the charge of an injected electron move
independently with
different velocities.
A more important result is the fact that
the quasiparticle weight $Z$  vanishes 
when the FS is approached.
This
 implies that for $k$ equal to the
Fermi momentum $k_F$, 
where spin and charge energies merge, the spectral function does
not become
 a delta function as a function of frequency $\om$, 
but rather it diverges with a weaker power-law behavior
  like $\om^{\al-1}$. This reflects onto the behavior of the momentum
 distribution $n(k)$, which no longer shows a discontinuity
 at $k=k_F$, but rather a
 power-law behavior ($|n(k) - n(k_F)| \propto |k-k_F|^\al$).
The same exponent appears in the local density of states, which
vanishes at $\om=0$ like $\om^{\al}$.
The exponent $\al$ thus characterizes the anomalous behavior of
one-particle correlation functions and it plays the role of the
anomalous dimension as in field theory. However, in contrast to the usual
field-theoretical
 models (as  $\phi^4$-theory), the anomalous behavior of
one-dimensional Fermions is not
universal, since the exponent $\al$ depends on the interaction.
1D metals having these properties take the name of Luttinger liquids
(LL), the name coming from 
 the Luttinger
model (LM)~\cite{lutt.63,ma.li.65,hald.81}, which plays the role of the
``canonical model'' for 1-D interacting fermions.

The interesting question is what happens between one  and two 
dimensions~\cite{cross.97,arri.99.c,ca.dc.me,bo.bo.95,ko.me.93,cl.st.94}.
Specifically, one can start from a $D$-dimensional array of chains
(the interesting cases are, of course,
$D=2$ or $3$),
initially uncoupled, and then switch on a small tunneling (hopping) 
amplitude \tp
between the chains. 
The question is when and how
does the crossover to a normal FL behavior occur?
While the question of the crossover from an anomalous LL to a normal
FL state is a challenging problem {\it per se},
there are  other reasons why one is interested in this problem.
The first two are  connected to the theory of high-\Tc superconductivity.
First, it has been suggested that
the normal-state properties of  high-\Tc
superconductors 
may be explained by some kind of two-dimensional LL
state~\cite{and.prl.90,and.sci.92}. Once a 2D LL state is
assumed within a CuO$_2$ plane, 
it has been suggested that incoherent hopping between different layers
may favor a BCS paired state~\cite{ch.an.94}.
Secondly, it has become clear from a variety of experiments\cite{tr.st.95}
that underdoped high-\Tc materials are characterized by the presence
 of charge modulations in the form of 
one--dimensional stripes~\cite{tr.st.95}.
In these  structures, the electron dynamics occurs mainly in
the direction longitudinal to the stripes, and, thus, it 
could be effectively described by quasi--one dimensional models
in which the transverse dynamics is reduced \cite{ca.ho.96,ca.gu.98}.
The
third reason is related to the existence of 
 several synthetic and natural 
compounds which can be considered 
as quasi-one dimensional metals~\cite{je.sc.82,gr.vo}, such as 
 the organic conductors
TTF-TCNQ, the Bechgaard salts~\cite{bech} $(TMTSF)_2X$ and
$(TMTTF)_2X$ (with $X=PF_6,\ ClO_4,\cdots$), or the inorganic chains
$NbSe_3$, $K_{0.3}MoO_3$.
A further possibility to study the crossover between 1D and
2D is to couple a finite number of chains together. The phase
diagram of such ladder systems is quite rich, and
it shows an interesting dependence 
on whether the number of chains is even or
odd
~\cite{da.ri.science,schu.96,ba.fi.96,3c.pla,or.gi.96,ar.br.96}.

In this paper,
 we consider
the effect of a small tunneling matrix element 
\tp coupling the 
 chains.
The question is: does the electron liquid go over to a FL state for
arbitrarily small \tp and sufficiently low temperatures or is there a
critical value of \tp below which one has a LL state for arbitrarily
low temperatures? This question is related to the problem of
dimensional coherence addressed by Anderson et al~\cite{and.sci.92,cl.st.94}.
These authors suggest that for sufficiently strong interaction
the system may remain in a LL state for sufficiently small \tp.
Clearly, the correct starting point, as stressed by these authors, is
to consider initially the problem of uncoupled LL and then treat \tp
as a perturbation.

However, renormalization-group 
calculations show that \tp is a relevant perturbation
which means that an arbitrarily small \tp should destroy the 1D LL
state~\cite{bo.ca.88}. 
This can be understood from simple dimensional arguments.
Consider
 the LL Green's function $\G(\xt|0)$ in real space~\cite{units}.
 This goes like $|\xt|^{-1-\al}$ at large distances,
 and thus the Fermi field operator $\Psi(\xt) \propto \sqrt{\G(\xt|0)}$
 has dimensions $[\Psi(\xt)]=\ww^{(1+\al)/2}$.
Therefore,upon integrating over the imaginary time $\tau$,
 the perturbation associated with the \tp term (see  
\eqref{htp} below), 
has dimension $\ww^{\al-1}$. 
This means that each term in the perturbation expansion in \tp carries
a term $\ww^{\al-1}$, which diverges at low energies whenever $\al<1$.
These divergences signal the fact that the perturbation \tp is relevant 
for $\al<1$.

Let us consider 
the energy at which higher-order terms in the \tp perturbation
start to become important (i. e., all of the same order). This is
given by  $\ww = \tp^{1/(1-\al)}\equiv \teff$. This introduces a new
 energy scale, $\teff$, which characterizes, for example, the
crossover  temperature above which temperature fluctuations cover the effect of
 \tp and the system behaves like a LL~\cite{bo.bo.95,compl}. 
This means that for temperatures $T$ much smaller than
$E_F$ but much larger than $\teff$ 
the scaling
 behavior is characterized by the LL anomalous dimension $\al$. For
 example, the Green's function at $k=k_F$ scales like
 $\omega^{\al-1}$ (for $\om\gg T$) in this range.
In this temperature region, the system is still effectively one
dimensional since the effects of \tp are washed out by the
temperature.
Below this crossover temperature~\cite{bo.bo.95}
 and for energies smaller than $\teff$
the effects of \tp become important and higher-dimensional coherence
sets on.
Notice that 
 the effect of electron interactions are indeed important
in reducing the coherence of the interchain hopping.
In fact, the crossover temperature is reduced considerably 
for $\al>0$, 
 since in this case $\teff\ll\tp$, and the interchain
hopping maintains an
 incoherent behavior down to very low temperatures~\cite{ko.me.93}.
However, strictly speaking, 
whether the system is a FL, a LL, or something else
 can  be determined in the $T\to
0$ limit only, since both  of them are {\it asymptotic} theories, i. e. valid in
the low-energy limit.
Therefore, the important energy region to be studied is
 $\ww \ll \teff$. This  is
 the nontrivial region, since the behavior here is determined by
all terms in the \tp expansion. 

For this reason,  
any  perturbative expansion restricted  to
lowest order  is  uncontrolled
at low energies $\ww\ll\teff$, and lowest-order expansions are inconclusive.
This is the reason why 
theoretical results
 are still contradictory 
about the nature of the ground state
in this energy region.
Since, as discussed above, this is precisely the 
relevant region for a possible FL behavior, it is worthwhile
investigating it in a controlled way.
This has been done in Ref.~\onlinecite{arri.99.c}, by considering all
diagrams corresponding to the infinite--dimension limit. 
In this paper, we extend the results of that Letter, and provide
the details of the calculation


This paper is organized as follows.
In Sec.~\ref{s:prob}, we introduce the problem of LLs weakly coupled by 
a single--particle hopping \tp. We discuss the issue of the
perturbation expansion in \tp, its difficulties, and the lowest--order
 approximations. Next, we discuss the limit considered here, namely the
``anisotropic'' $D\to \infty$ limit, and the analysis of the
asymptotic low--energy regime.
Finally, we present an appealing discussion of the
 analogy of our method  with the parquet summation and with the renormalization
 group, and discuss the  cases in which the present method is controlled.
In Sec.~\ref{s:dinf}, we describe in detail 
 the procedure to
 carry out the sum of the diagrams leading to
the $D\to\infty$ 
limit for the self energy \eqref{gamma}. 
 The idea is to write a recursive equation for the
``restricted renormalized cumulants'' \eqref{gc} in terms of the
effective hopping $\td$. In the leading logarithmic order, this gives
a set of self--consistent recursive equations, \eqref{ff0}, which can be easily
solved to a very high degree of accuracy 
by a power expansion and a Pad\'e analysis.
In Sec.~\ref{s:resu}, we discuss the results of this calculation. The
most important one is the fact that the anomalous exponent scales to zero, 
i. e., the self energy no longer scales anomalously at low energies.
This is seen in the spectral function close to the ``special'' Fermi
point $\cp=0$, which becomes sharper, in contrast to the 
 the lowest--order approximation. The quasiparticle
weight no longer vanishes at $\cp=0$ in our result.
Finally, in Sec.~\ref{s:conc} we state our conclusions,  and discuss
possible extensions of the calculation to the inclusion of
spin--charge separation and to 
finite dimensions.

Due to the absolute novelty of our procedure, we 
considered that the reader would benefit from an inclusion of
all details of the calculations, 
so that any one could follow 
and repeat our steps without  difficulties, and possibly extend them to
some other cases.
The calculation is transparent, as it
 is almost completely analytic 
except for the Pad\'e 
solution of the
recursive equation described in Sec.~\ref{s:solu}.
In order not to burden the bulk of the
paper,
we deferred
 most of these calculational details to the
appendices.

\section{The problem: from one to higher dimensions}
\label{s:prob}

 We consider a $D'$ ($=D-1$)-dimensional hypercubic array of
parallel one-dimensional  chains (i. e., the total dimension is $D$).
We consider here the case of equal spin and charge velocities,
 since it allows for crucial simplifications in the calculation.
Since we are interested in the effects and in the 
fate of the anomalous exponent $\al$, 
we believe that 
spin-charge separation should not play an important role. 
The chains are labeled by the 
 $D-1$-dimensional coordinate $\xy$ along the 
 hyperplane perpendicular to them, while
the coordinate along the chains is called $\xx$. 
The  Hamiltonian we want to study has the following form~\cite{units}:
\beq
\label{htp}
H= \sum_{\xy} H_{LL}(\xy) +  \sum_{\xy \xy'} \tp(\xy-\xy')\ 
  \sum_{r\g} \int d \xx \ 
\psi^{\dag}_{r,\g}(\xx,\xy) \ \psi_{r,\g}(\xx,\xy')\;,
\eeq
where 
$\psi_{r,\g}(\xx,\xy)$ [$\psi_{r,\g}^{\dag}(\xx,\xy)$] 
is the destruction [creation] operator for a right- ($r=+1$) or
left-moving ($r=-1$) fermion 
at the position $\xx$ along the chain $\xy$
with spin $\g$.
Moreover,
$H_{LL}(\xy)$ is the  Hamiltonian for an
(uncoupled)  LL in the 
chain $\xy$. Since we are interested in low--energy properties
we can just take for  $H_{LL}(\xy)$ a Luttinger model,  characterized 
by its parameters $\al$ and
$v_F$~\cite{lutt.63,ma.li.65,voit.rev} (since we neglect spin--charge
separation), 
which  will depend in a
nontrivial way on the bare parameters of the microscopic chain 
Hamiltonian. 
However, we are not interested in this dependence here, and we
just take these parameters as our starting point.
 In \eqref{htp}, $\tp(\xy-\xy')$
is the amplitude for the hopping of an electron
from chain $\xy'$ to chain $\xy$, where, as usual, we have
assumed that  neither the $\xx$
coordinate, nor the direction $r$ are changed by the hopping. 
Moreover, one can restrict to the case of an hopping between
nearest--neighbor chains only.
Inclusion 
of an hopping with 
 finite extension in the $\xx$ direction, or of a next--nearest--neighbor
hopping  in the $\xy$ direction 
is straightforward. However, it is 
 not expected to change the low-energy results.
With $\tp=0$, the problem can be solved exactly, as the
ground state is  given
by the product ground states of 
the LM in each chain, which is known~\cite{lutt.63,ma.li.65,hald.81}.

Knowing the exact solution of the $\tp=0$ problem,
one can envisage
 carrying out a perturbative expansion
in powers of \tp, as \tp is small.
 This is, however, not without complications, as 
Wick's theorem does not hold for the 
$\tp=0$ ground state, since the LM, although exactly solvable,
 contains  electron--electron interactions.
A similar problem occurs for the expansion about the atomic limit of the
Hubbard model, whereby one first solves the single--site problem exactly
and then expands in powers of the hopping $t$. 
 A diagrammatic formulation for this problem
was introduced by Metzner in
Ref.~\onlinecite{metz.91}, and further discussed in Ref.~\onlinecite{pa.se.98}.
  It consists in carrying out a linked--cluster expansion, where an
 arbitrary (even) number of lines ($2n$) can 
 join into one dot. This dot is associated 
with  the exact $n$-particle cumulant 
of the single-site problem~\cite{cumul}.

 This method has been
extended to the problem of expanding about the LLs in Ref.~\onlinecite{cross.97}.
The diagrams contributing to the expansion are the same, the only
difference being that each line is now labeled by the extra variable
$\xx$ (intrachain coordinate) and $r$ (for left- or right-moving fermions),
besides spin $\g$, and imaginary time $\tau$.
Actually, this method turns out to be more appropriate for the present
problem rather than for the Hubbard model. Indeed, in the Hubbard
model,
one expands about an highly degenerate $t=0$ ground state, which is not the
case in our problem of coupled LLs at $\tp=0$.
Alternatively, one can use the diagrammatic rules in momentum space,
for which each line carries an intrachain momentum $\kx$, a
Matsubara frequency $\om$, and an interchain momentum $\ky$, as well
as  indices $\g$ and $r$. 
Besides this modification, rule 2 of Ref.~\onlinecite{metz.91} for
calculating the Green's function remains the same.
A set of these curious diagrams, contributing to the Green's function,
are shown in 
 Fig.~\ref{diag}.
The building blocks of the diagrammatic expansion are
(i) hopping lines connecting 
nearest-neighbor chains (say $\xy_1,\xy_2$)
 associated with  $\tp(\xy_1-\xy_2)$, and 
(ii) ``dots'' with $n$ entering and $n$ leaving legs,
associated with the $n$-particle cumulant of the single chain.
The latter 
can be readily evaluated,
 at least for low energies, since one knows the exact solution of the 
Luttinger model and of its correlation functions (cf. Sec.~\ref{s:corr}). 
\levfig{
\begin{figure}
   \centerline{
\psfig{file=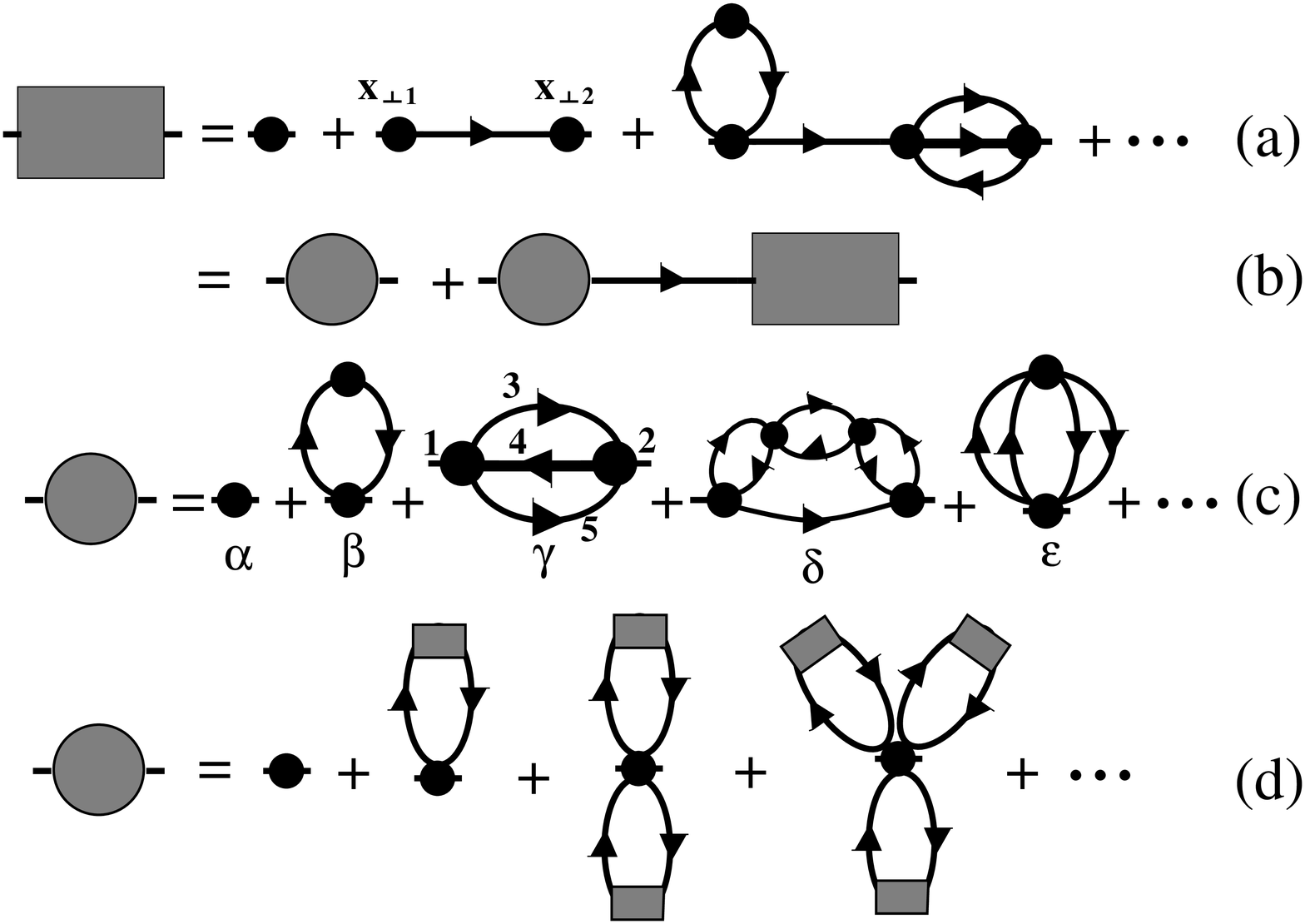,width=\figwidth}
}
\vspace*{.3cm}
\caption{
\label{diag}
Diagrammatic expansion in $\tp$ of the single-particle
  Green's function $\G$ (gray box). A directed line connecting two
  chains
$\xy_1$ and $\xy_2$
 gives a contribution
  $\tp(\xy_1-\xy_2)$, or $\tp(\ky)$ in momentum
  space~\cite{units}.
A dot with $n$ entering and $n$ leaving lines   contributes a factor 
$\G^0_c$
($n$-particle 
 cumulant of the uncoupled LL, see Sec.~\ref{s:dinf}).
 (a) Example of single-particle irreducible and
  reducible contributions to $\G$. (b) Dyson's equation for $\G$ in
  terms of the inverse-self-energy $\Gamma$ (gray disk). 
(c) Example of diagrams  contributing to $\Gamma$.
(d) Self-consistent diagrams contributing to $\Gamma$ in the
 $D\to\infty$ limit. The self consistency is due to the presence 
 of the full $\G$ in the internal lines of the loop.
}
 \end{figure}
}

Boies et al. used a functional--integral method 
to obtain  an expansion in \tp about the LL~\cite{bo.bo.95}.
Although their formulation allows, in principle, for an expansion
to any order in \tp, in practice one can just get the first few
orders.
Our method provides a systematic diagrammatic formulation of this
expansion to any order. 
The advantage of a diagrammatic formulation is that one can 
choose a class of diagrams to sum over, according to some physical
guidance, without being restricted to the few lowest--order
terms. This is particularly important for the model at study, since,
as discussed in the Introduction,
each power of \tp in 
the perturbation 
carries
a term $\ww^{\al-1}$, which diverges precisely in the important
region. Thus, one cannot reliably restrict to a finite number of diagrams.

Some diagrams contributing to the
 expansion of the Green's function $\G$ (gray box) 
are shown in Fig.~\ref{diag}a. As in conventional perturbation
theory, one can consider the function $\Gamma$ obtained by 
the sum of  irreducible diagrams, i. e., the ones which
 cannot be separated by cutting a
single line (see Fig.~\ref{diag}c).
One then obtains a Dyson-like equation for $\G$ as a function of $\Gamma$
(Fig.~\ref{diag}b) of the form~\cite{units}
\beq
\label{dyson}
\G(\kom) = \left[\Gamma(\kom)^{-1} - \tp(\ky)\right]^{-1} \;.
\eeq
Notice that $\Gamma^{-1}$,
and not $\Gamma$,
appears in the inverse Green's function
 contrary to  standard perturbation theory. 
For this reason, we call $\Gamma$
inverse self energy.

The lowest--order approximation for $\Gamma$ 
(the ``dot'':
 $\al$ in Fig.~\ref{diag})
corresponds to
taking $\Gamma=\G^0$,  the Green's function of the isolated LL.
 This gives
 for the total Green's function \eqref{dyson}~\cite{units}
\beq
\label{gwen}
\G(\kom) = \left[\G^0(\kom)^{-1}-\tp\cp\right]^{-1} \;.
\eeq
This expression is a generalization of
 the Hubbard I approximation for the case of an expansion about the LL.
\eqref{gwen} has been first obtained by Wen via a different
procedure~\cite{wen.90}, and reobtained by  Boies et
al.~\cite{bo.bo.95} within a functional--integral method.
 This approximation, which we will refer to as ``\LO'',
 is also called
``single--dot'', ``RPA'',  ``Wen's'', or ``Hubbard I'' in other papers. 
For $\al<1$, 
the effect of the interchain kinetic energy 
$\tp\cp$ is to change the branch-cut
singularity into a true quasiparticle pole (cf. Ref.~\onlinecite{scho.cc.97}) 
for all $k$ points close to the FS, except
for those $\ky$ points for which $\cp=0$ (for example, for $D=2$ these 
are $\ky=\pm \pi/2$).
In particular, the positions of the poles for $\om=0$ identify the new
FS, which acquires a dispersion of the form $\kx_F(\ky) \propto (\tp
\cp)^{1/(1-\al)}$, i. e. it is reduced with respect to the
noninteracting case, where one would have $\kx_F(\ky) \propto \tp \cp$, but
not completely suppressed\cite{cross.97}.
For the sake of completeness, we discuss the main results of this
approximation in Sec.~\ref{s:lo}.

Since the branch cut are shifted into poles,
 this approximation gives a FL along the whole FS except close
to the $\cp=0$ region. This can be also seen from the quasiparticle
weight $Z$, plotted in Fig.\ref{figfigz} (dashed line), 
which vanishes for $\cp=0$. For this reason, the quasiparticle peak is
quite broad in this region, as can be seen from Fig.\ref{dispwen}.
However, as  discussed above, this result, being restricted to lowest
order is uncontrolled in the region $\ww\ll\teff$ and one should sum
an infinite series of diagrams in order to get reliable results. 
Since it is not possible to sum 
all diagrams in the expansion, we want to select a
workable subset of diagrams according to some {\it physical} limit in order to
avoid an arbitrary choice.
Specifically, we consider the series given by the diagrams
indicated in Fig.~\ref{diag}d, corresponding to the large-dimension
limit ($D\to\infty$).
The $D\to\infty$ procedure adopted here 
is different from the standard dynamical
mean-field theory ~\cite{dinf}, since
our system is strongly anisotropic, as
the hopping in one (in the $\para$) direction is not rescaled by the
usual $1/\resc$ factor and is  much larger than in
the other $D-1$ ($\perp$) directions~\cite{units}.
In analogy to the standard $D\to\infty$ method~\cite{dinf}, 
where one has a single 
{\it impurity} embedded in a self--consistent medium,
our $D\to\infty$ system represents a
{\it 1-D chain} embedded in an effective self-consistent medium. 
As a consequence, the self-energy is local with respect to
the $\perp$ coordinates 
but has a nontrivial dependence on 
the $\para$ ones~\cite{local}.
We believe that this is the correct starting point to study the
crossover problem, since, in this way, one treats the
one-dimensional problem exactly and includes the coupling to the other 
chains by
an effective dynamical mean field.

Even summing all the $D=\infty$ diagrams is an impossible
task. Nevertheless, since we are interested in low-energy properties,
we can restrict to the leading singularities in each diagram.
It turns out  convenient to 
rewrite the power expansion in terms of the 
{\it dressed} hopping \td (indicated by a dashed line in
Fig.~\ref{diagdinf}). This is very similar to the skeleton expansion
in conventional perturbation theory, where self-energy insertions are removed. 
The advantage is that the scaling behavior of the effective hopping
(cf. \eqref{tdx0}) exactly cancels the power-law 
divergences of the diagrams,
and 
each term of the perturbation acquires the same
scaling as a function of the energy, and 
 only logarithmic divergences are left.

The  procedure of summing just the leading logarithmic divergences 
 is similar in spirit to the sum of the leading
divergences in the parquet series, which was introduced by the
Russian~\cite{russian} and by the French~\cite{french} school in order
to study the instabilities of various
one- and higher-dimensional electron systems. 
This method is equivalent to the one-loop renormalization-group ($g$-ology) 
approach\cite{soly.79}, and it actually
gives a rigorous background, as well as a systematic formulation for
the extension of the $g$-ology method to higher dimensions.
In our case,
 this corresponds to considering the quantity 
$l=\al \log( \teff/\ww)$ to be of order one, and thus 
 taking all orders in $l$, while considering $\al$ small. 

Similarly, in the parquet summation, or $g$-ology\cite{soly.79}, 
the small parameter is the bare interaction vertex $g_0$ and one 
sums all powers of
$g_0 \log \frac{E_F}{\om}$, $\om$ being the characteristic  energy scale. 
The sum of this series gives the {\it renormalized} interaction
vertex 
$g(\om/E_F)$ which thus acquires  an energy dependence.
Within the renormalization--group picture, the energy-dependent 
interaction vertex is interpreted as 
an effective interaction acting on an effective low-energy subspace,
i. e., on a subspace in which high-energy modes are integrated out.
Whenever the
 interaction vertex scales to zero, this signals  that the effective
 low-energy theory   describes non-interacting electrons,
 i. e., the theory is asymptotically (infrared) free. As a
 consequence, the exponents of correlations functions are mean-field
 like and, in the case of fermions, the system is a Fermi liquid.
On the other hand, when a vertex diverges, no controlled 
prediction can be made about the low-energy behavior of the system,
since the perturbative approach breaks down for sufficiently low
energies, even when $g_0$ is small. 
In this case, the divergent vertex signals an instability
towards some kind of broken-symmetry state.

In our case, the role of the interaction vertex is played by the
anomalous exponent $\al$.
The {\it bare} $\al$ is the correlation exponent of the uncoupled set
of Luttinger liquids. Switching on the interliquid hopping \tp produces
a renormalization of the exponent. This renormalized exponent is
obtained  by looking at the low-energy behavior of the
self-energy  in the coupled-chains system. 
Similarly to the $g$-ology case,
our result,  obtained by summing the leading logarithmic divergences,
 is thus controlled if (i) the starting (bare) value of
$\al$ is not too large and (ii) $\al$ scales to zero for low energies.
The first (i) requirement is easy to fulfill, since for most interesting
systems $\al$ is quite small
For example, for the Hubbard model $\al
\leq \frac18$, where the equal sign holds for an infinite value of 
the on-site interaction 
$U$. Larger
values of $\al$ are obtained by increasing the range of the
interaction \cite{ed.97}. This is another reason why our approach is more
convenient than a weak-coupling expansion in  $U$: while our
calculation makes sense also for very large (bare) $U$, for which $\al$ is
still small, the weak-coupling renormalization group
 is not justified for $U$ larger
than the bandwidth.
An estimate of the maximum value $\al_c$ of $\al$, 
for which our calculation is justified is given in Sec.~\ref{s:conc}.
 The second (ii) requirement can be only checked {\it a
  posteriori}.
The main result of this paper is that indeed
point (ii) turns out to be  satisfied, as
$\al$ scales to zero for energies smaller than \teff.
Thus, our procedure of restricting to the leading logarithmic
divergences is controlled, unless one starts from a model with a too
large value of $\al$.
\levfig{
\begin{figure}
   \centerline{\psfig{file=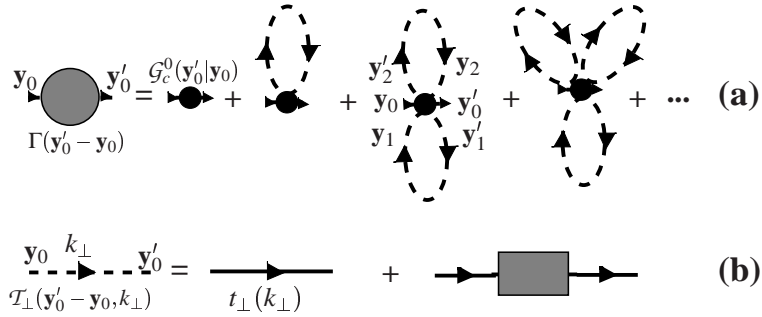,width=\figwidth}}
\vspace*{.3cm}
\caption{
\label{diagdinf}
(a) Diagrams contributing to the inverse-self-energy $\Gamma$ 
 in the
       $D=\infty$ limit within an expansion in the dressed hopping \td
       (dashed line). 
(b)  Dressed hopping  and its diagrammatic expression in
terms of the bare hopping \tp (full line) and the Green's function.
Other conventions are as in Fig.~\ref{diag}.
}
 \end{figure}
}

\section{Anisotropic $D \to \infty$ method}
\label{s:dinf}

In this section, we carry out the sum of the $D\to\infty$ diagrams for 
the inverse-self-energy.
In
 the $D\to\infty$ limit, 
the inverse-self-energy
$\Gamma(\xt_0)$ 
is $\perp$-local~\cite{local}, and 
is obtained as the sum
of the loop diagrams 
in Fig.~\ref{diagdinf}a (equivalent to the ones of Fig.~\ref{diag}d) as
\beqn
\label{gamma}
\nonumber && \Gamma(\xt_0) = 
\G^0_c(\xt_0|0) + \sum_{m=1}^{\infty}
\frac{(-1)^m}{m!}
\int
\Bigl[ \prod_{k=1}^m
d^2 \yt_k \ 
d^2 \xt_k \ 
\td(-\xt_k,0) \Bigr] \, 
\G^0_c(\yt_0+\xt_0, \cdots, \yt_m+\xt_m | \yt_{0}, \cdots, \yt_{m}) \;,
\\ &&  =
\G^0_c(\xt_0|0) + \sum_{m=1}^{\infty}
(-1)^m
\int_{1\reg m} 
\Bigl[ \prod_{k=1}^m
d^2 \yt_k \ 
d^2 \xt_k \ 
\td(-\xt_k,0) \Bigr] \, 
\G^0_c(\yt_0+\xt_0, \cdots, \yt_m+\xt_m | \yt_{0}, \cdots, \yt_{m}) \;,
\eeqn
where in the last line we have exploited the symmetry for exchange of the coordinates
$1,\cdots, m$ and restricted the integration
to the region
$|\xt_1| > |\xt_2| > \cdots > |\xt_m|$ indicated by
 ``$1\reg m$''.
 The corresponding factor  $m!$
is then canceled by the symmetry factor  $1/m!$ of the diagram.
In  \eqref{gamma},
$\G^0_c(\yt_0' \cdots \yt_m' | \yt_{0} \cdots \xt_{m})$ 
is 
the $m+1$-particle cumulant of the uncoupled LL, 
i. e. the connected part of the
$m+1$-particle Green's function
$\G^0_c(\yt_0' \cdots \yt_m'| \yt_{0} \cdots \xt_{m})$ defined in
\eqref{g0ph} (see also \eqref{cumdef} for the definition of cumulants
in terms of Green's functions).
In particular, for $m=0$ the single--particle cumulant $\G^0_c$
coincides with the Green's function $\G^0$, as there are no
disconnected parts.
Moreover,
 $\td(\xt,x_{\perp}=0)$
is the dressed  hopping written in real space, which is calculated in Sec.~\ref{dressed}.

We are interested in the dominant
low-energy 
behavior
($\ww\ll \teff$ corresponding to 
$|\xt_0| \teff \gg 1$)
of  correlation functions and thus we can restrict to
the leading logarithmic divergences in  the loop
integrals (\eqref{gamma}), as discussed in Sec.~\ref{s:prob}.
Let us estimate this leading contribution. If, as a first step, one neglects
the self-consistency of the Green's function and
 dresses the
hopping $\td$ with the bare Green's function only (\eqref{tdx0}), one
can see that the leading contribution of 
a $m$-loop term in \eqref{gamma} has the form $\G^0_c(\xt_0|0) \times (\al \log
|\xt_0| \teff)^{2m}$. Indeed, one ``$\al \log$'' term arises from each
integration of the ``center of mass'' coordinates $\yt_k$
(cf. Sec. \ref{intcent}), another ``$\al$'' from each $\td$, due to
its 
 real-space structure
 (cf. \eqref{tdx0}), and a ``$\log$'' comes out for each integration
of the ``relative'' coordinates (\eqref{ffm}).

Even summing up ``just'' the leading logarithmic divergences  of the
integrals in \eqref{gamma} is a tough task. To do this we proceed in
several steps. 
First, consider that some integration regions in \eqref{gamma} can be
left out, as they don't contribute to the leading logarithmic
divergences.
Specifically,
in addition to the region $|\xt_1| > |\xt_2| > \cdots >
|\xt_m|$ (called $1\reg m$), to which we restrict by symmetry, we
 can further restrict to the region\cite{great}
$|\xt_0| > |\xt_1|$, and
$|\xt_p| < min(|\yt_q-\yt_r|,|\yt_q'-\yt_r|,|\yt_q'-\yt_r'|)$
for each $p \geq q,r $  (of course, $q\not=r$), and $\yt_q'$ is
defined as $\yt_q+\xt_q$.
The fact that the leading logarithmic contributions only come from 
this integration region, which we will call
``$0\Reg m$'',
is proven in Sec.~\ref{proofrelevant}.

For convenience, we introduce the ``restricted renormalized
cumulants'' (RRC)~\cite{rrc}, defined only in the region
 ``$0\Reg m$'' as
\beqn
\label{gc}
&&
\G_c(\yt_0+\xt_0, \cdots, \yt_m+\xt_m | \yt_{0}, \cdots, \yt_{m})
\\ \nonumber &&
\equiv
\G^0_c(\yt_0+\xt_0, \cdots, \yt_m+\xt_m | \yt_{0}, \cdots, \yt_{m})
\\ \nonumber && 
-
\int_{0\Reg m+1}
d^2 \xt_{m+1}  \ 
d^2 \yt_{m+1}  \ 
\td(-\xt_{m+1},0) 
\\ \nonumber && \times
\G_c(\yt_0+\xt_0, \cdots, \yt_{m+1}+\xt_{m+1} | \yt_{0}, \cdots,
\yt_{m+1}) \;.
\eeqn
Comparing \eqref{gamma} and \eqref{gc}, 
it is straightforward to verify that 
$\Gamma(\xt_0)$. is given by
the single-particle RRC, 
$\G_c(\xt_0|0)$.

We thus proceed by evaluating the integrals in \eqref{gc}.
An important point, which we will show below, is that,
 at the leading logarithmic order, the $m+1$-particle cumulant is
renormalized by a multiplicative factor, which depends on
the absolute values of the relative coordinates $|\xt_i|$ only.
More precisely, 
 the RRC can be written as
\beqn
\label{gcff}
&&
 \G_c(\yt_0+\xt_0, \cdots, \yt_m+\xt_m | \yt_{0}, \cdots, \yt_{m})
\\ && \nonumber
=
 \FF_m(l_{0},\cdots,l_{m}) \ 
 \G^0_c(\yt_0+\xt_0, \cdots, \yt_m+\xt_m | \yt_{0}, \cdots, \yt_{m})
\eeqn
where the $\FF_m$ is the  renormalization factor, which we have
 written in terms of the logarithmic variables
 $l_{i}\equiv \al \log(|\xt_i| \teff)$. 
We can thus first carry out
the integration over the ``center-of-mass'' coordinate $\yt_{m+1}$  in
 \eqref{gc} by simply considering the effect on the 
 bare cumulant, as the renormalization factor 
does not depend on 
$\yt_{m+1}$.
This integral is quite involved, but its leading logarithmic
 contribution can be calculated analytically.
This is carried out in Sec.~\ref{intcent}, where one obtains~\cite{units}
\beqn
\label{inty}
&& \int\limits_{0\Reg  m+1} d^2 \yt_{m+1} 
\; \G^0_c(\yt_0+\xt_0, \cdots, \yt_{m+1}+\xt_{m+1} | \yt_{0}, \cdots, \yt_{m+1}) 
\\ &&  \nonumber
=
2 \pi \
\G^0_c(\yt_0+\xt_0, \cdots, \yt_m +\xt_m| \yt_{0}, \cdots, \yt_{m}) 
\\ && \nonumber
\times
 |\xt_{m+1}|^2 \ 
\G^0_c(\xt_{m+1}|0) \ 
\sum_{j=1}^m \al \ 
\left(\log|\xt_j| - \log|\xt_{m+1}|\right)\ 
\left(\delta_{r_j,r_{m+1}} + \frac{1}{S}\ \delta_{r_j,-r_{m+1}} \right)
 \;.
\eeqn

After carrying out the integral over 
$\yt_{m+1}$ we carry out the
integration over the ``relative'' coordinate $\xt_{m+1}$, which
includes a sum over $r$ and $\g$~\cite{units}.
Inserting the form \eqref{gcff} and  the result \eqref{inty} into
\eqref{gc}, and dividing both sides of the equation by
$\G^0_c(\yt_0+\xt_0, \cdots, \yt_m+\xt_m | \yt_{0}, \cdots, \yt_{m})$, 
one obtains
\beqn
\label{ffm}
&&
 \FF_m(l_{0},\cdots,l_{m}) = 
1 - 2 \   \pi \ \int\limits_{\teff^{-1}<|\xt_{m+1}|<|\xt_m|} d^2
\xt_{m+1} \ \td(-\xt_{m+1},0) \ 
 |\xt_{m+1}|^2 \ 
\G^0_c(\xt_{m+1}|0) \  
\\ \nonumber &&
\times 
\FF_{m+1}(l_{0},\cdots,l_{m},l_{m+1}) \ 
\sum_{j=1}^m 
\left(l_j-l_{m+1}\right) 
\ \left(\delta_{r_j,r_{m+1}} + \frac{1}{S}\ \delta_{r_j,-r_{m+1}} \right)
\;,
\eeqn
where the lower limit
 of integration for $|\xt_{m+1}|$ is due to the fact that
$\td$ changes its behavior
in the region $ E_F^{-1} < |\xt_{m+1}|
< \teff^{-1}$ (Ref.~\onlinecite{tdnolog})
, and, thus,   
there is no logarithmic contribution here,
 and the upper one is due to the restriction
$0\Reg  m+1$
  in \eqref{gc}. 
Inserting the asymptotic expression for the dressed hopping \eqref{tdx1}
in \eqref{ffm}, one can carry out
the integration over $\xt_{m+1}$ 
in circular coordinates, and
 obtain the recursive
 self-consistent equation 
for $\FF_m$
\beqn
\label{ff0}
&&
\FF_m(l_0,\cdots,l_m) = 1+ 2(1+S)  \ \int_0^{l_m} d l_{m+1} \ 
\sum_{j=0}^m  \ (l_j-l_{m+1})
\\ && \nonumber
\times \ 
\FF_{m+1}(l_0,\cdots,l_{m+1}) \ 
\left[\bar \FF_0(l_{m+1}) + \bar \FF_0'(l_{m+1}) \right] \;.
\eeqn
From \eqref{ff0}, it is obvious that 
$\FF_m$ depends on just two variables, namely
$l\equiv l_0+\cdots+l_{m-1}$, and $l_m$. With this redefinition,
 and renaming the integration variable $l_{m+1}$ to $l'$, \eqref{ff0} can be reduced to
\beq
\label{ff}
\FF_m(l,l_m) = 1+ 2(1+S)  \ \int_0^{l_m} d l'\ 
 [l+l_m- (m+1) \ l']\  
\FF_{m+1}(l+l_m,l')\ 
\left[\bar \FF_0(l') + \bar \FF_0'(l') \right] \;.
\eeq

\eqref{ff} is a self-consistent equation, since
 $\bar \FF_0 = 1/\FF_{m=0}$ [\eqref{bfr}], which depends on the $\FF_m$, 
to insert on the r.h.s..
We have not been able to find an analytic solution to \eqref{ff}.
However, by expanding in powers of 
of the variables $l_i$ one can write a recursive equation for the
coefficients of the expansion of the functions $\FF_m$ up to a
rather high order with a moderate numerical  effort. 
This procedure is described in detail in Sec.~\ref{s:solu}.

We have evaluated the coefficients of $\FF_0$ up to the 
$42^{th}$ order in $l$.
A Mathematica program has allowed us to evaluate these coefficients in a
rational form, which is particularly recommended for a Pad\'e analysis. 
A straight summation of 
the series is not recommended, since its convergence radius seems
to be rather small (of the order unity), while we need the
asymptotic behavior for large $l$. 
Nevertheless $l=\al\log(|\xt|\teff)$ is restricted to the neighborhood
of the real positive axis, 
and  a Pad\'e analysis shows that the
poles are either away on the complex plane or on the negative real axis.
A Pad\'e analysis is thus the most appropriate procedure in order to determine
the large-$l$ behavior of the function $\FF_0(l)$, which also gives
the asymptotic behavior of the inverse self-energy $\Gamma(x)$.
The results will be presented and discussed in Sec.~\ref{s:resu}.

\section{Results and discussion}
\label{s:resu}

As shown in Sec.~\ref{s:solu}, 
the solution of \eqref{ff} gives 
 $\FF_0(l) \sim  e^{c l}$ for large $l$, where the exponent $c$ turns out to be
essentially equal to $1$ (within about $ 10^{-4}$ of accuracy) in both 
cases with and without spin.
Introducing this result and \eqref{g0ll} in the expression for
the inverse-self-energy
(\eqref{gcff} with $m=0$)
yields
\beq
\label{gamf0}
\Gamma(\xt) = \G^0(\xt|0) \ \FF_0(\al \log (|\xt| \teff)) \to 
\G^0(\xt|0) \ (|\xt|\ \teff)^\al \propto \teff^{\al}/|\xt| \;,
\eeq
i. e. the anomalous 
exponent $\al$ exactly cancels out in the asymptotic behavior of $\Gamma$!.
The same thing happens in momentum space. 
From \eqref{gamk1} one notices that 
 the (asymptotic behavior of the) renormalization function is 
the same in momentum space, provided one 
replaces $|\xt|$ with $1/|\kom|$.
Thus, for low energies we obtain
for the right-moving component ($r=+1$)
\beq
\label{gammares}
\Gamma(\kom)  
\to \G^0(\kom) \ \teff^{\al} \ |\kom|^{-\al}
\propto \teff^{\al} \ (\iom-\kx)^{-1} \;,
\eeq
where we have used \eqref{g0k}.

The Green's function of the coupled system 
is given by the Dyson equation \eqref{dyson}.
Taking the result \eqref{gammares}, one can readily notice
 that  the
Green's function now has  poles at $\iom - \kx \propto \teff^{\al}
 \tp(\ky)$, i. e. 
even for $\cp=0$, in contrast to the \LO result, where a branch
cut was present.
In particular, 
at the FS ($\iom\to 0 + i 0^+$) and for $\cp=0$, our result becomes
asymptotically exact, as $|\kom|$ vanishes at the pole.

Let us look at the FS more precisely.
This  is the curve
$\kx_F(\cp)$
 parametrized
by the Fermi momentum as a function of the $\perp$ momenta, and is
determined by  the solution of the equation 
$\Gamma(\kx_F(\cp),\iom=0+i 0^+)^{-1}= \tp \cp$. 
Obviously, \eqref{gammares} gives $\kx_F(\cp=0)=0$. In
Fig.~\ref{figfigfs} we plot the FS curve for  other values of $\cp$,
and $\al=1/4$ in the case of particles with spin.
 We compare our result (full line)
  with  the \LO result (dashed line).
For small \cp, our result gives a regular behavior
 $\kx_F(\cp) \propto \teff \ \cp$ in contrast to  the lowest-order
 result, which gives a flattening of the FS at $\cp=0$, due to the behavior 
$\kx_F(\cp) \propto \teff
\  \cp^{1/(1-\al)}$. 

The quasiparticle weight $Z(\cp)$ at the FS  is given by the inverse
of the coefficient of the linear term in $\om$ in the inverse Green's
function, more precisely,
$Z(\cp)^{-1} =  \frac{d}{d i \om} 1/\G(\kx_F(\cp),\iom)_{\iom\to0+i
  0^+}$.
We have plotted $Z$ as a function of $\cp$ for the case with spin in 
 Fig.~\ref{figfigz}, again compared with the \LO
 approximation.
Moreover, in order to show the importance of summing the {\it infinite} 
series of diagrams, we have included the 
result obtained by truncating the $D\to\infty$ series
(Fig.~\ref{diagdinf}a) at the first loop, by still
 taking the
self--consistently dressed hopping as internal line. 
For small \cp, the lowest-order result (dashed line)  gives a
$Z$ vanishing as $Z(\cp) \propto (\tp \cp)^{\al/(1-\al)}$, 
 thus yielding poorly defined 
quasiparticles
around
$\cp=0$. Inclusion of the 
first loop
 (dotted line) gives a vanishing Z too. Therefore, self consistency is 
 not enough to restore the FL behavior.
 Our result, instead,
yields a finite $Z$ for $\cp\to0$, as can be seen from the figure
(solid line).
The correct FL behavior is thus recovered 
on the {\it whole } FS, including the regions $\cp=0$.

\levfig{
\begin{figure}
   \centerline{\psfig{file=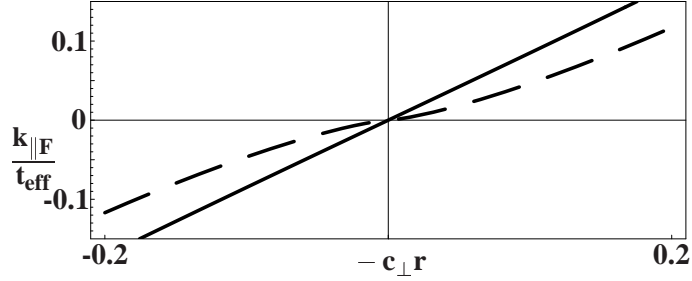,width=\figwidth}}
\caption{
\label{figfigfs}
Fermi-surface dispersion $\kx_F$ 
as a function of the off-chain kinetic energy \cp 
 (in units of \tp)  for the coupled spinful Luttinger
 liquids [\eqref{htp}] with bare LL 
 exponent \mbox{$\al=1/4$}.
Our $D\to \infty$ result (solid line)
is compared
 with the \LO approximation \eqref{gwen} (dashed).
}
 \end{figure}
}

\levfig{
\begin{figure}
   \centerline{\psfig{file=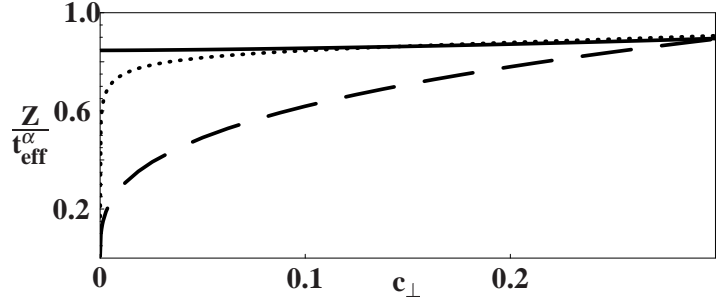,width=\figwidth}}
\caption{
\label{figfigz}
Quasiparticle weight Z 
as a function of the off-chain kinetic energy \cp with the same 
conventions as in 
 Fig.~\ref{figfigfs} 
. 
In addition, we show the result 
(dotted line)
obtained by partially improving on the \LO approximation, i. e., by
including 
 the first self-consistent loop  for the inverse
 self-energy of 
 Fig.~\ref{diag}d.
}
 \end{figure}
}

These results can be more concretely seen in the
spectral function
for small $\cp$~\cite{ancont}. This is plotted in Fig.~\ref{dispdinf}
for different $\cp$, and for $\kx=0$.
The figure shows a well--defined dispersive quasiparticle peak, which becomes 
sharper by approaching the FS, as should be the case for a FL.
The dispersion as a function of $\cp$ is a clear indication of
higher--dimensional coherence.
For comparison, in Fig.~\ref{dispwen}, we have shown the \LO
result.
As one can see, the peak is dispersive too, but much 
broader and lower (notice the different scale). Moreover, a closer
inspection shows that
 the quasiparticle weight decreases by
approaching the FS, which is consistent with what
 we have shown in Fig.~\ref{figfigz}.

\levfig{
\begin{figure}
   \centerline{\psfig{file=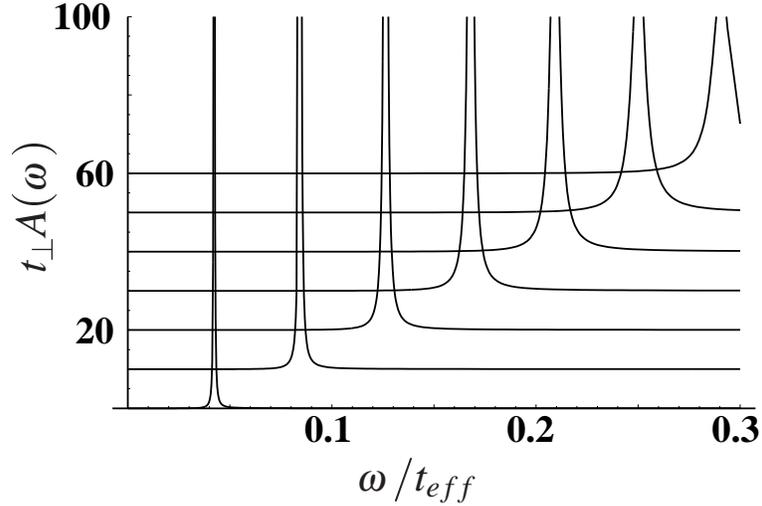,width=\figwidth}}
\caption{
\label{dispdinf}
Spectral function $A(\om)$ 
of the 
 coupled spinful Luttinger liquids
for different values of $\cp$,  $\kx=0$,
and $\al=1/4$
from our $D\to \infty$ result.
For the sake of clarity, 
the different curves 
are  shifted vertically by steps of $10$.  
They correspond to 
 $\cp=0.05, 0.1, 0.15, 0.2, 0.25, 0.3 ,0.35$, from bottom to top.
Notice the sharpening of the peaks upon approaching the FS at $\cp\to 0$.
 }
 \end{figure}
}

\levfig{
\begin{figure}
   \centerline{\psfig{file=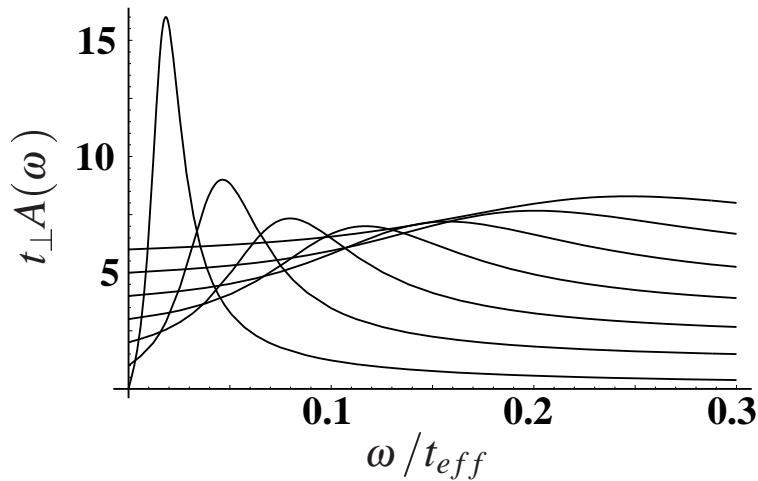,width=\figwidth}}
\caption{
\label{dispwen}
Spectral function $A(\om)$ 
from the \LO approximation 
for the same parameters as in Fig.~\ref{dispdinf}, except that
 the curves are shifted by $1$.
The peaks sharpen upon approaching the FS, but the quasiparticle
 weight vanishes.
}
 \end{figure}
}

We want to study the spectral function even for $\kx\not=0$. To
understand what happens, let us first look at the spectral function 
for the LM~\cite{me.sc.92,voit.93} (without spin-charge separation),
 which we plot in Fig.~\ref{llspec} for $\kx=0.2$.
From the figure, one can readily recognize the two nonanalicities at 
$\om\pm \kx$. For $\om\searrow +\kx$ one has in fact a
divergence like $(\om-\kx)^{\al/2-1}$, while for $\om\nearrow -\kx$ the
spectral function vanishes as $(\kx+\om)^{\al/2}$. 
The power--law divergence instead of a pole at $\om=\kx$ is due to the
fact that the point,
$\om=\kx$,
where the inverse Green's function, $1/{\G^0}$, of the LL  vanishes,
 is not a simple zero but a branch cut.
Between $\pm \kx$ the spectral function of the LM is identically zero, 
as the Green's function has neither cuts nor poles here.
At $\kx=0$ the two nonanalicities merge in a single power-law
divergence $\om^{\al-1}$.

\levfig{
\begin{figure}
   \centerline{\psfig{file=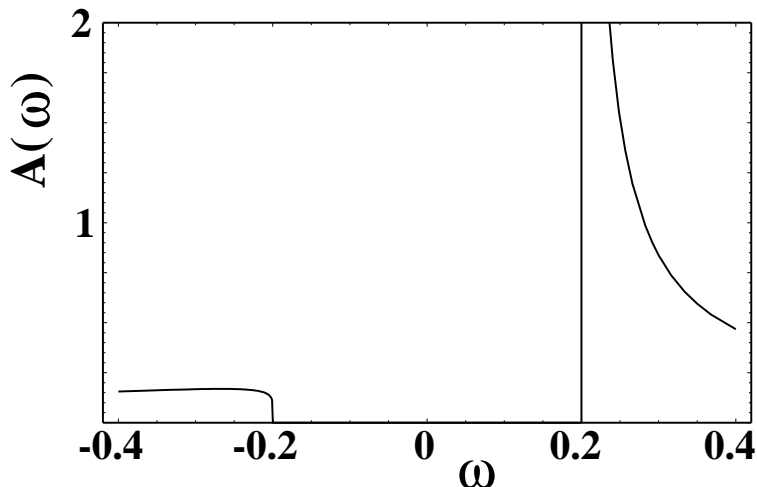,width=\figwidth}}
\caption{
\label{llspec}
Spectral function of the isolated Luttinger model (with equal charge
and spin velocities) for
$\al=1/4$ and $\kx=0.2$~\cite{me.sc.92,voit.93}.
}
 \end{figure}
}

\levfig{
\begin{figure}
   \centerline{\psfig{file=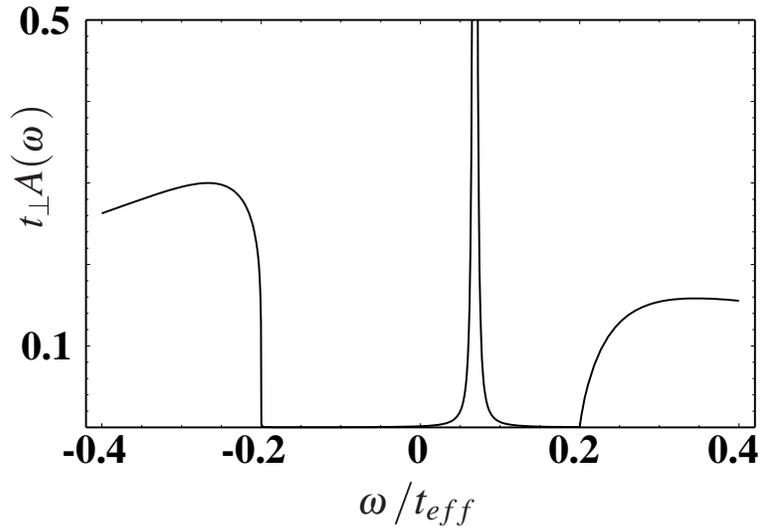,width=\figwidth}}
\caption{
\label{polewen}
Spectral function 
within the \LO approximation for 
the coupled spinful Luttinger liquids with
$\al=1/4$, $\kx=0.2$, and $\cp=-0.2$.
In order to make the quasiparticle delta function visible, we have
added a small imaginary part $\sim 3.0 \ 10^{-5}$. Due to the proximity
of the singularity, the peak actually becomes  broader.
}
 \end{figure}
}

\levfig{
\begin{figure}
   \centerline{\psfig{file=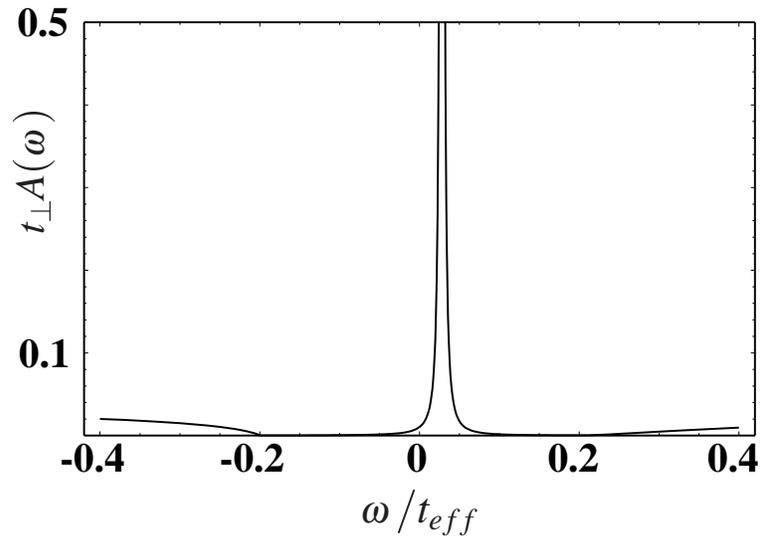,width=\figwidth}}
\caption{
\label{poledinf}
Spectral function 
for the $D \to \infty$ result
with the same parameters as in Fig.~\ref{polewen}.
Notice the much larger transfer of spectral weight from the
singularities to the quasiparticle pole.
 }
 \end{figure}
}

Within the \LO approximation, \eqref{gwen}, the zero of $1/{\G}$ is
shifted away from the branch cut. Thus, an isolated quasiparticle 
pole appears in
the region $-\kx<\om<\kx $ on the real axis. 
This pole is always present for any
$\cp\not=0$ (see Sec.~\ref{s:lo}).
The pole removes spectral weight from the peak at $\om=\kx$
which is  now no longer a divergence.

In our $D\to \infty$ result, the situation is similar. However,
Fig.~\ref{poledinf} shows that  in
this case the two singularities at $\pm \kx$ 
loose much more spectral weight in favor of the pole.
 This is another  reason
why the quasiparticle weight remains larger within our result, as 
 shown in Figs.~\ref{figfigz}, and \ref{dispdinf}.
For $\cp=0$ \eqref{gwen} does not have quasiparticle poles,
while our result yields a
pole with nonvanishing weight at the FS even in this case.
The reason for that is due to the different behavior of the
quasiparticle weight, as shown in Fig.~\ref{figfigz}, and by
the fact that the scattering rate does not vanish fast
enough for \eqref{gwen}, while it vanishes faster than linearly within 
our result, as discussed in
Ref.~\onlinecite{arri.99.c}  
(cf. Fig.~2c of that reference).

\section{Conclusions}
\label{s:conc}

In conclusion, we have studied the problem of the 
crossover from one 
to higher dimensions for fermionic systems, 
when  Luttinger liquids are coupled by a small
hopping \tp.
Specifically, we have concentrated on the region below the
single--particle crossover temperature, $\ww\ll\teff$,
which is the one relevant for the dimensional crossover.
We have carried out an expansion in powers of \tp, and summed
  the self--consistent series of diagrams (Fig.~\ref{diag}d)
  corresponding to the {\it anisotropic} $D\to \infty$ limit.
Our result shows that 
 the LL exponent $\al$ renormalizes to zero 
for energies smaller than the single-particle crossover temperature
$\teff$. The system thus
flows to a FL fixed point with mean-field like
 exponents.
This is seen, for example, in the self--energy, which now scales
linearly as a function of frequency and momentum, in contrast to the
\LO approximation \eqref{gwen}, where the self--energy still scales
anomalously like
$|\kom|^{1-\al}$ .
As a consequence,
well defined quasiparticles are recovered along and in the
neighborhood of  
the whole FS, in contrast  to the result \eqref{gwen}, where the spectrum is
incoherent for small \cp.

We have shown the importance of including an infinite series of
diagrams, in order to give reliable results in
the region $\ww\ll\teff$. Even introducing the first loop of the
diagrammatic expansion (in Fig.~\ref{diag}d) does not give the
correct result, as shown in Fig.~\ref{figfigz}. This shows 
  that not even  a  self--consistent 
 calculation is  sufficient. This is the reason why previous 
theoretical results, restricted to lowest orders,
 are still contradictory 
about the nature of the ground state
in this energy region.

These results 
have been 
obtained for the case of equal spin and charge velocities.
In fact,
we believe that the scaling behavior
of the anomalous exponent $\al$ found here
is universal and should not be affected by the inclusion of
spin--charge separation. 
Nevertheless,
an extension of the present calculation to
the case of LLs with different velocities
 could be interesting, first, in order 
 to check this fact, and second, 
in order to verify whether spin--charge separation scales as well to
zero
in higher dimension like $\al$, or not.

The imaginary part of the self energy, $-Im \ \Gamma^{-1}$,
 needed to evaluate the spectral 
functions in Figs.~\ref{dispdinf} and \ref{poledinf} has been
determined by analytic continuation of the {\it asymptotic } form
\eqref{gammares}~\cite{ancont}.
However, 
 one should mention that our calculation, restricted to the leading
divergences, 
yields reliable results for  $Im \Gamma^{-1}$ 
at  small values of $\om$ and $\kx$ only.
On the FS and for large \cp,  $\kx$ is large too.
Thus, for large \cp, we cannot state with certainty whether
corrections 
to $Im \Gamma^{-1}$
beyond the leading divergences  
vanish fast enough upon approaching the FS or not.
Arguments similar to the one 
of ordinary perturbation theory\cite{lutt.61}
cannot be extended to the present case,
due to the
momentum dependence of the vertices in the \tp expansion.
A hint can be possibly obtained by explicitly evaluating numerically 
the first few
loops in Fig.~\ref{diag}d, without restricting to the leading divergences.

In principle, we cannot say
whether our result is valid also for the physical cases of finite
dimensions, and, in particular for 
$D=2$ or $D=3$. 
However, as we have shown in Sec.~\ref{s:nonl},
 the non-$\perp$-local dressed hopping 
$\td(\xt,x_{\perp}\not=0)$  vanishes 
faster than
 the $\perp$-local one $\td(\xt,0)$  for large $|\xt|$.
Non $\perp$-local contributions are thus irrelevant and one may try to 
extend the present
result  to finite dimensions.
However, there are still 
{\it $\perp$-local} diagrams of order $1/D$, (for example,
if one takes the diagram $\gamma$ of Fig.~\ref{diag} and replaces
all internal line with a local $\td$), which may spoil this result.
It might be interesting to consider an expansion about the present
$D=\infty$ result, and consider the irrelevance or relevance of such
diagrams, and give predictions about a possible critical dimension
$D_c$, above which the results of this paper hold.
For example, this could be done in order to
study the critical behavior in the 
 neighborhood of the transition 
to the two--particle regime at $\al=\al_{2p}$\cite{cross.97},where
$\al_{sp} \sim 0.41$ ($0.62$) for spinless (spinful) electrons.

In Sec.~\ref{s:prob},
we have already discussed 
that our ``renormalization--group--like'' 
result holds for $\al$ smaller than a certain
$\al_c$. 
Although
we cannot determine $\al_c$ exactly within our approach, we
can estimate it, e. g., by the value of $\al$ 
for which the spectral function becomes
negative in some regions. This criterion gives $\al_c \approx 0.50$ for 
the spinless and $\al_c\approx 0.33$ for the spinful case.

Another question is the contribution of the shifted poles
$|\kom|,\cp\not =0$,
 which turn out 
 to be irrelevant in
the present case (cf. Sec.~\ref{dressed}). 
However, these poles may give  important contributions in lower
dimensions. Indeed, these poles are the one giving rise, in some
conditions, to the well known  nesting or
superconducting instabilities at selected regions of the FS.

The author thanks W. Hanke for useful discussions.
Partial support by the BMBF (05SB8WWA1) is acknowledged.

\appendix

\section{Many--particle correlation functions of the Luttinger
  liquid}
\label{s:corr}

For the sake of completeness, and in order to fix our notation,
 we give here the expressions
for the $n$--particle Green's functions of the LM in real space.
To our knowledge, their explicit expression, although known, 
has not been reported anywhere else.
In Sec.~\ref{s:scal}, we
 discuss the scaling behavior of the diagrams in the \tp
expansion.

A generic
$\perp$-local $n$-particle  Green's function  is defined as~\cite{units}
\beq
\label{ggen} 
 \G^{d_1 \cdots d_{2n}} (\xt_1,\cdots,\xt_{2n})
\equiv
 < T_{\tau} \psi^{d_1}(\xt_{1})  \cdots  
\psi^{d_{2n}}(\xt_{2n}) > \;,
\eeq
where 
$T_{\tau}$ is the imaginary-time ordering operator,
$ \psi^d(\xt) $  destroys (for $d=-1$) or creates 
(for $d=+1$)
a fermion at the point $\xt$  (which includes $r$ and $\g$).
In order to extract the $\tp=0$ cumulants 
$\G^0_c$ of the isolated LM, to be used in \eqref{gamma}, we
first need
the (disconnected) Green's functions $\G^0$.
These
can be written as
\beq
\label{gll}
 \G^{0,d_1 \cdots d_{2n}} (\xt_1,\cdots,\xt_{2n}) = (2 \pi a)^{-n}
\eta_{r_1} \cdots \eta_{r_{2n}} \ 
\prod_{2n \geq i_1>i_2 \geq 0} 
\left[P_{r_{i_1} r_{i_2}}(\xt_{i_1}
-\xt_{i_2})\right]^{-d_{i_1} d_{i_2}} \;.
\eeq
This holds
whenever 
the  particle- and momentum-conservation constraints
$\sum_{i=1}^{2n} d_i=0$, and  $\sum_{i=1}^{2n} d_i r_i=0$
 are fulfilled, otherwise $\G^0=0$.
Here, $a$ is a short-distance cutoff ($a \propto \frac{v_F}{E_F}$).
The Klein factors $\eta_{r_i}$  
obey  anticommutation rules
$\{\eta_{r},\eta_{r'}\} = \delta_{r,r'}$ and  
account for the fermionic anticommutations \cite{schu.cll.96,ba.ho.76}.
From now on,
we will set $a$ to
unity, unless otherwise specified.

The functions $P$ in \eqref{gll} can be  written as
\beq
\label{P}
P_{r_1 r_2}(\xt) =
 \R(\xt)^{- q(r_1\ r_2,\QQ)} \ 
e^{i  \frac{r_1+r_2}{2}\  \left(\frac{\pi}{2} - A(\xt)\right)}
\;,
\eeq
where the exponent $q(r_1,r_2,\QQ)$ is given by~\cite{units}
\beqn
\label{pexp}
q(r,\QQ) = && \left\{ \matrix{ 
    \frac{1}{2}\ (r \QQ + \frac{1}{\QQ}) & \hbox{ for } S=1 \cr
    \frac{1}{4}\ [r (\QQ+1) + \frac{1}{\QQ}+1] & \hbox{ for } S=2
 }\right. 
\\ \nonumber &&
= 
\left\{ \matrix{ 1+ \al & \hbox{ for } r=1 \cr
            B & \hbox{ for } r=-1 \cr
 }\right. 
\;,
\eeqn
where the LL exponent $\al$ 
is related to  $\QQ$ via
\beq
\label{Qal}
\frac{\QQ+1/\QQ-2}{2\ S}=\al \;,
\eeq
and
\beq
\label{defB}
B\equiv \frac{1/\QQ-\QQ}{2 S}\;.
\eeq
Here,
\beq
\label{rxt}
\R(\xt) \equiv \frac{\be}{\pi a}\  \sqrt{\cosh^2 \widetilde{\xx} -
  \cos^2 \widetilde{\tau}} \;,
\eeq
and
\beq
\label{axt}
A(\xt) \equiv \arg(\tanh \widetilde{\xx} + i \tan \widetilde{\tau}) \;,
\eeq
$\widetilde \xx \equiv \frac{\pi \xx}{\be}$,
$\widetilde \tau \equiv \frac{\pi \tau}{\be}$.
At zero temperature $T=\frac{1}{\be}=0$, \eqref{rxt} and \eqref{axt}
become
\beq
\label{rxt0}
\R(\xt) \to \sqrt{\frac{\xx^2+\tau^2}{a^2}} = \frac{|\xt|}{a}\;,
\eeq
and
\beq
\label{axt0}
e^{i A(\xt)} \to e^{i \arg(\xx+i \tau)} = \frac{\xx+ i \tau}{|\xt|} =
\frac{\xt \cdot \vv}{|\xt|} \;,
\eeq
where we have introduced the complex vector
 $\vv=(1,i)$, allowing for a compact expression.
These expressions are valid for $|\xt|\gg a$ and need a short-distance
cutoff for $|\xt|\sim a$. The 
 cutoff prescription for the LM
 amounts to replacing
$ \xx^2+\tau^2$ with $\xx^2+(|\tau|+a)^2$. However, it turns out
convenient to
 adopt a ``rotation symmetric'' cutoff
 obtained by replacing
$ \xx^2+\tau^2$ with $\xx^2+\tau^2+a^2$, or by setting 
$\R(\xt)=1$ for $|\xt|<a$.
 The low-energy results, obviously, don't
depend on the specific choice of the short-distance cutoff.
The advantage of setting equal spin and charge velocities is clear 
at this point. Without this assumption, the correlation functions
would  not be invariant under rotation in the $(\xx,\tau)$
plane, which would have made the  calculations more difficult.

In conformity with Ref.~\onlinecite{arri.99.c}, we define
\beqn
\label{g0ph}
 \G^0(\yt_0', \cdots, \yt_m' | \yt_{0}, \cdots, \yt_{m})
&&\equiv
\G^{0,1,-1,1,-1,\cdots}(\yt_{0},\yt_{0}',\cdots,\yt_{m},\yt_{m}') 
\\ \nonumber &&
=
 < T_{\tau} \psi^{\dag}(\yt_{0}) \psi(\yt_{0}')  \cdots  
\psi^{\dag}(\yt_{m}) \psi(\yt_{m}') >_{\tp=0} \;,
\eeqn
and 
$ \G_c^0(\yt_0', \cdots, \yt_m' | \yt_{0}, \cdots, \yt_{m})$ as the
corresponding cumulant (or connected Green's function) to be inserted
in the diagrammatic expression \eqref{gamma}.

As an example, we use \eqref{gll} to evaluate the single- and the
two-particle Green's functions (here, we indicate
explicitly the indices $r$ as $1$ for $r=+1$ or $\bar 1$ for $r=-1$).
The single-particle Green's function reads 
\beq
\label{g0ll}
\G^0(\xt \, 1|0 \, 1) = -\frac{i}{2 \pi a} \ |\xt|^{-1-\al}\ e^{- i \arg
  \xt \cdot \vv} \;,
\eeq
while the two-particle Green's function for right-moving particles
reads
\beq
\G^0(\yt_1  1,\yt_2  1|\yt_1'  1,\yt_2'  1) = 
  \frac{\G^0(\yt_1  1|\yt_1'  1) \ \G^0(\yt_2  1|\yt_2'  1) \ \G^0(\yt_1
     1|\yt_2'  1) \ 
    \G^0(\yt_1'  1|\yt_2  1)}{ \G^0(\yt_1  1|\yt_2  1)\  \G^0(\yt_1'
     1|\yt_2'  1)} \;.
\eeq
On the other hand, the two-particle Green's function for mixed right-
and left-moving particles reads
\beq
\label{gmixed}
\G^0(\yt_1 1,\yt_2 \bar1|\yt_1' 1,\yt_2' \bar1) = 
 \G^0(\yt_1 1|\yt_1' 1) \ \G^0(\yt_2  \bar1|\yt_2' \bar1) \ 
\left[ \frac{ |\yt_1-\yt_2'|\  |\yt_1'-\yt_2|}{|\yt_1-\yt_2| \ |\yt_1'-\yt_2'|}
\right]^{-B} \;.
\eeq

\subsection{Scaling behavior of diagrams}
\label{s:scal}

From Eqs.~(\ref{g0ph},\ref{gll},\ref{P},\ref{rxt0},\ref{Qal})
 one can easily extract the scaling behavior of Green's
functions for a homogeneous rescaling of the coordinates 
$\xt_i\to \lam \xt_i$.
\beq
\label{scal}
 \G^0(\lam \yt_0', \cdots, \lam \yt_{n-1}' | \lam \yt_{0}, \cdots, \lam \yt_{n-1})
= \lam^{-(m+1)\ (1+\al)}
 \G^0(\yt_0', \cdots, \yt_{n-1}' | \yt_{0}, \cdots, \yt_{n-1}) \;,
\eeq
i. e.,
 an $n$-particle Green's function (and a cumulant too) 
scales like $n$ one-particle
Green's functions in real space.
Going back to the diagrammatic formalism, 
\eqref{scal} gives  the scaling behavior of a vertex with $2n$ legs.
In addition, each internal line, associated with a \tp term,
 contributes an integration over $\tau$ and $\xx$,
i. e. a factor $\lam^2$.
Let us now consider a
order-$N$ diagram ($N$ internal lines), with $E$ external
lines. Each internal line belongs to two vertices, while each external 
one to one, so that the 
sum over all  vertices ($v$) of the number of legs for each vertex
$L_v$ is equal to
$L\equiv  \sum_v L_v = 2 N+E $. Adding the contribution from the
integrals in the internal lines, this diagram scales like
$ \lam^{- (1+\al)\ L/2 +2 N} =\lam^{(1-\al)\ N- (1+\al)E/2}$.
This shows that each order in $\tp$ contributes a factor
$\lam^{1-\al} \sim \ww^{\al-1}$ (Ref.~\onlinecite{units}).
To get
the same diagram in momentum space one has to  integrate over $E-1$
external $\xx$ and $\tau$,  getting a factor $\lam^{2E-2}$. For
example, a momentum--space
diagram of order $\tp^N$ for the inverse self--energy 
scales like $\ww^{-(N+1)(1-\al)}$.

This is correct provided
no short--distance divergences occur in the
integration of diagrams, i. e. if the integrals do not depend on the
short--distance cutoff $a$ of \eqref{rxt0}.
A short--distance divergence would introduce a negative power of
$a$, which has to be compensated by a positive power of $\lam$
in order to have the correct dimensions (powers of a length
scale). 
 This is what happens,
e. g. in diagrams $\gamma$ and $\delta$ in Fig.~\ref{diag}c, for
$\al>\al_{2p}$, i. e. in the two--particle regime\cite{cross.97}. 
In diagram $\gamma$, if one assigns
 to the external lines ($1,2$) the index $r=+1$,
and to the 
 internal lines $3,4,5$ the
 indices $r=+1,-1,-1$, respectively,
and inserts the expressions for the two two--particle vertices taken from
\eqref{gmixed}, one obtains for $\Gamma(\xt_1-\xt_2)$
a contribution of the form (we don't consider
the dependence on the $\perp$ coordinate here)
\beqn
\label{scal2p}
\nonumber
&&  a^{4\al} \ \tp^3 \   \int 
\prod_{i=3}^5 d^2 \xt_i \  
\left( \fc{\xt_1-\xt_3} \fc{\xt_4-\xt_5} \right)^{-1-\al} 
\\  &&\times
\left[ 
\left(\frac{\fc{\xt_1-\xt_5} \fc{\xt_3-\xt_4}} {\fc{\xt_1-\xt_4} \fc{\xt_3-\xt_5}}\right)^{-B} 
-1 \right] 
\\ \nonumber &&\times
 e^{-i (\arg(\xt_1-\xt_3) - \arg(\xt_4-\xt_5))}
\\  \nonumber &&\times
\left[ 
\left(\frac{\fc{\xt_2-\xt_5} \fc{\xt_3-\xt_4}} {\fc{\xt_2-\xt_4} \fc{\xt_3-\xt_5}}\right)^{-B} 
-1 \right] 
\\ \nonumber &&\times
 e^{-i (\arg(\xt_3-\xt_2) - \arg(\xt_5-\xt_4))} \;.
\eeqn 
According to 
the scaling analysis   carried out above,
 the contribution \eqref{scal2p} should behave like
$ a^{4 \al} \ \tp^3  \ |\xt_1-\xt_2|^{2-4\al}$ (notice that
this expression correctly has the dimensions of an inverse length).
This behavior is correct by assuming that the integral does not depend 
on $a$ in the $a\to 0$ limit.
However, this is not the case for $B>1$, for which
the integral diverges at small distances, as one can readily verify.
Thus,  for $B>1$ the integral gives
an $a$-dependent
contribution $a^{2-2B}$, 
which must be balanced by  an additional contribution 
$\propto |\xt_1-\xt_2|^{2B-2}$ in order to have
the correct dimensions. 
Thus, for $B>1$, corresponding to
 $\al>\al_{2p}$, the contribution
\eqref{scal2p} goes like
$ \tp^3 \ a^{2-2B + 4 \al} \  |\xt_1-\xt_2|^{2 B -4\al}$, i. e. a
stronger divergence.
This produces the two--particle exponent obtained in Ref.~\onlinecite{cross.97}.

\section{Results of the lowest--order approximation}
\label{s:lo}

In this section, we summarize some results of the LO approximation
\eqref{gwen}
introduced by Wen\cite{wen.90}. 
Within  this  approximation, 
the introduction of \tp 
modifies the denominator of the Green's function by a term $\tp\cp$.
 The Green's function for the LM \eqref{g0k} 
can be readily analytically continued to the complex plane
(we set the
constant $g_{\al}$ to $1$ for simplicity, and take $r=+1$): 
\beq
\G^0(\kx,z=\iom) = \frac{(\om^2+\kx^2)^{\al/2}}{\iom-\kx} =
 \frac{(\kx^2-z^2)^{\al/2}}{z-\kx} \;.
\eeq
This expression is analytic for $z$ on the real axis and $-\kx < z <\kx$, which
is the reason why the LM spectral function is zero in this region.
The denominator of  \eqref{gwen} becomes
\beq
\label{denom}
\G^0(\kx,z)^{-1} - \cp \tp = \frac{z-\kx}{(\kx^2-z^2)^{\al/2}} - \cp
\tp \;.
\eeq  
The zero of \eqref{denom} gives  a true pole
whenever it occurs within the region of analicity.
For example, the FS is given by the points $\kx, \, \cp$ where
\beq
 \G^0(\kx,z=0)^{-1} = \cp \tp \;,
\eeq
i. e.
\beq
\label{polefs}
-\kx/|\kx|^{\al} = \cp \tp \Rightarrow \kx = -\sign \cp \ |\cp
\tp|^{1/(1-\al)} \;.
\eeq
By including a finite value for the energy $z$, one can easily see that, whenever
$\cp\not=0$,
 \eqref{denom}
is analytic in a neighborhood
 of this point, i. e., the solution is a true 
pole (cf. Ref.~\onlinecite{scho.cc.97}).
 By differentiating \eqref{denom} with respect to $z$ 
and replacing the solution \eqref{polefs}, one obtains the inverse of
the residuum, i. e. of the weight $Z$, for this pole. 
The result is
$Z=
(\cp \tp)^{\al/(1-\al)}$, and is plotted in 
Fig.~\ref{figfigz}. 

Close to  the FS, one
can look for a zero of \eqref{denom} of the form $z= x \ \kx$. This
gives
\beq
\kx = y(x) \ \kx_F(\cp) 
\eeq
with
\beq
y(x) = 
 \left(
  \frac{(1+x)^{\al/2}}{(1-x)^{1-\al/2}}\right)^{1/(1-\al)} \;.
\eeq
The solution is real, and thus it gives a pole, for each $-1<x<1$.
In this region, $y(x)$ takes all the values $0<y(x)<\infty$, i. e.,
for each $y>0$ there is always a solution $x$. This
 means that 
for any point ($\kx,\cp$) in the Brillouin zone with $\kx \cp<0 $ 
one always has a pole at a given frequency.
The weight $Z$ of the pole is readily evaluated as
\beq
Z= |\cp \tp|^{\al/(1-\al)} \left( (1-x)^{-\al/2} (1+x)^{1-\al/2}
\right)^{1/(1-\al)} (1+(1-\al) x)^{-1} \;,
\eeq 
which vanishes only at the border of the region, $x \to \pm 1$.
Obviously , the above discussion 
only holds for $\al<1$.

The fact that there is always  a true pole for any $\kx \cp<0 $ 
can be also seen  directly from \eqref{denom}. For given $\cp \tp$ 
(say $>0$), and $q=-\kx>0$, the function
\beq
\frac{q+z}{(q^2-z^2)^{\al/2}}
\eeq
vanishes for $z=-q$ and diverges for $z\to q^-$. Between $-q$ and $+q$, 
it is an increasing function of $z$. 
Thus, for any $\tp \cp$, there
is always  a value of $z$ within the analytic region of \eqref{denom},
giving a zero. In practice, for small $\cp$ the pole starts to build
close to the left nonanalicity, while for increasing $\cp$ it
approaches the right singularity. This pole can be seen in
Fig.~\ref{polewen}.

\section{Evaluation of the dressed hopping}
\label{dressed}

In this section, we evaluate the long--distance behavior of the
dressed hopping in real space, which we need in \eqref{ffm}.
Its diagrammatic  equation  is given 
in Fig.~\ref{diagdinf}b and reads\cite{units}
\beq
\label{td}
\td(\kom,\cp) = \tp \cp  + \tp \cp\ \G(\kom,\cp)\ \tp \cp
= \tp \cp\ ( 1- \tp \cp \ \Gamma(\kom))^{-1} \;,
\eeq
where we have used the Dyson equation \eqref{dyson}.
At the lowest order, $\Gamma$ scales as
$\Gamma(\kom)\sim \ww^{\al-1}$,  and thus,
from \eqref{td}, $\td(\kom,\cp)$ formally goes like $\ww^{1-\al}$ for
small energies and fixed $\tp \cp$.
As discussed in the Introduction (cf. also Sec.~\ref{s:corr}),
 every order in \tp in the perturbation
expansion carries along a term which scales like $\ww^{\al-1}$ and
thus higher orders in \tp are stronger and stronger divergent.
However, due to the scaling of \td,
replacing
 the bare hopping \tp  in the perturbation expansion 
with the dressed \td cancels this power-law divergence. 
Thus,
the correct
starting point  to study the low-energy region is  to
carry out a {\it skeleton} expansion in \td  and remove all
self-energy insertions. In our case, this corresponds to replacing the 
diagrammatic series of Fig.~\ref{diag}d with the one of
 Fig.~\ref{diagdinf}a.
 Although the power-law singularities have
disappeared in this way, logarithmic divergences are still present in
this expansion as discussed in Sec.~\ref{s:dinf}.
 These divergences can, however,
be resummed, in the same spirit as it was done for
the parquet series by Dzyaloshinskii and by Nozi{\`e}res and coworkers
\cite{russian,french}, as discussed in Sec.~\ref{s:prob}

The behavior of \td discussed above 
holds for nonzero $\cp$. In \eqref{ffm} we need the $\xy=0$ 
hopping, i. e. we have to integrate over $\cp$, including $\cp=0$.
One should, thus, treat this integration point with due care.
We first Fourier transform in the $\perp$ direction, obtaining
\beq
\label{tdxy0}
\td(\kom,\xy=0)= \int d \cp \ \D(\cp) \td(\kom,\cp) \;,
\eeq
 where $\D(\cp)$ is the density of states for the out--of chain energy. 
In the $D\to\infty$ limit and for a cubic lattice
with nearest--neighbor hopping this
reads~\cite{units,dinf}
\beq
\label{dens}
\D(\cp) = \frac{1}{2 \sqrt{\pi}} e^{-\cp^2/4}\;.
\eeq
The integral \eqref{tdxy0} can be readily evaluated for small
energies, where the quantity $\eps\equiv\left[\tp
  \Gamma(\kom)\right]^{-1}$
 is small.
By inserting \eqref{td}, 
collecting $\eps $
and summing and subtracting $\tp$
 yields
\beq
\label{tdxy01}
\td(\kom,\xy=0)= - \eps \ 
 \int d \cp \ \D(\cp)  \left( \tp + \frac{\eps\   \tp}{\cp-\eps} 
\right) 
= -\eps \ \tp \left[ 1 + O(\eps \ \log \eps) \right] \to -
\frac{1}{\Gamma(\kom)} \;,
\eeq
where the $\log \eps$ contribution is given by the $\cp=0$ point.
It is clear that 
the asymptotic result \eqref{tdxy01} does
not depend on the specific form of the density
of states $\D(\cp)$, as long as it is regular at $\cp=0$ and normalized.

We now carry out the Fourier transform in the $\para$ direction.
For the sake of definiteness, we consider here the 
 right--moving ($r=+1$) component.
All results for $r=-1$ are simply obtained by changing the sign of
the $\para$ coordinate, i. e. $x_{\para}$ or $k_{\para}$.
As a first go, we evaluate $\td^0$, 
the \LO approximation
for \td, i. e., we use 
$\Gamma=\G^0$. The full dressed \td will be evaluated in 
 Sec.~\ref{s:full}.
The LL Green's function is given in \eqref{g0ll}.
Its Fourier transform  is given by
\beq
\label{g0k1}
\G^0(\kom) = \int d^2\xt\  e^{-i \kom \cdot \xt}\  \G^0(\xt|0) \;,
\eeq
where we have identified~\cite{units} $\kom=(\kx,-\om)$.
 We now
introduce the angles of the two vectors $\xt$ and $\kom$ with
the ``$x$'' axis, i. e. 
 $\phi = \arg \xt \cdot \vv$, and $\th = \arg \kom \cdot \vv$, and
 $\vv=(1,i)$ as  in Sec.~\ref{s:corr}.
 \eqref{g0k1} becomes
\beq
\G^0(\kom) = \frac{- i a^{\al}}{2 \pi} \ \int d^2\xt \ e^{-i |\xt||\kom|
  \cos(\th-\phi)}\ e^{-i \phi}\ |\xt|^{-1-\al} \;.
\eeq
Going over to
circular coordinates and transforming
$\phi'=\phi-\th$, and $s=|\kom||\xt|$ yields
\beq
\label{g0k2}
\G^0(\kom) = \frac{- i a^{\al}}{2 \pi}\ |\kom|^{\al-1}\ e^{-i \th}
\int_{|\kom| a}^{\infty} s^{-\al} d s \ \int_0^{2 \pi} d \phi'\
e^{-i(\phi'+s \cos\phi')} \;,
\eeq
where we don't care about the specific form of the cutoff at
 $s<|\kom| a$, as it can be taken to zero
in the
low-energy limit ($|\kom|$ is always limited by $\teff\ll 1/a$). 
The last
 integral over $\phi'$ gives $- 2 \pi i J_1(s)$, with $J_1(s)$ a
Bessel function. 
Integrating over $s$,  \eqref{g0k2}
gives
\beq
\label{g0k}
\G^0(\kom) \approx - \frac{|a\kom|^{\al}}{\kom \cdot \vv}\ g_{\al} 
= - \frac{|a\kom|^{\al}}{\kx - i \om}\ g_{\al} 
\;,
\eeq
where
\beq
g_{\al} \equiv \frac{\Gamma(1-\al/2)}{2^{\al} \Gamma(1+\al/2)} \;.
\eeq
Like in \eqref{g0k}, we will from now on 
  indicate with ``$\approx$'' expressions valid in the asymptotic
  limit.
However, whenever 
this will become clear, we will switch back to ``$=$''.

To evaluate
$\td^0(\xt,0)$, we first insert \eqref{g0k} 
in \eqref{tdxy01}
(remember, here we use $\Gamma=\G^0$), and then transform  
back into $\xt$ coordinates.
Thus,
\beqn
\label{tdun}
&&
\td^0(\xt,0) \approx - \int \frac{d^2 \kom}{4 \pi^2}\ e^{i \kom \cdot \xt}\
 \G^0(\kom)^{-1}
\\ \nonumber &&
 = \frac{1}{4 \pi^2 g_{\al} a^{\al}}\ \int q \ d q\ d \th\ e^{i q |\xt|
   \cos(\th-\phi)}\
q^{1-\al}\   e^{i \th} \;,
\eeqn
with the same conventions as above, 
and with $q\equiv|\kom|$.
Transforming $q |\xt|=s$ and integrating over $\th$, yields
\beq
\label{tdun1}
\td^0(\xt,0) \approx 
\frac{|\xt|^{\al-3}\ e^{i \phi}}{4 \pi^2\ g_{\al}\ a^{\al}}
\int_0^{\infty} s^{2-\al}\  d s \ 2 \pi i\  J_1(s) \;.
\eeq
In principle,
the last integral does not converge
 at large $s$. However, it
can be regularized by inserting a convergence factor $e^{-\mu s} =
e^{-\mu |\xt| |\kom|}$ with $\mu \sim \frac{1}{|\xt| \teff}$, physically
due
by the fact
that
the behavior  $\td(\kom,\cp) \approx - \Gamma(\kom)^{-1} $
[\eqref{tdxy01}] is cutoffed
at $|\kom| \sim \teff$. The convergence factor $\mu$ can then be safely 
taken to zero, since the result of the integral does not depend on
$\mu$ for small $\mu$.
In this way, one obtains
\beq
\label{tdx0}
\td^0(\xt,0) \approx 
\frac{i \al\ (2-\al)}{2 \pi a^{\al}}\ 
|\xt|^{\al-3}\ e^{i \phi} \;,
\eeq
with $\phi=\arg \xt\cdot \vv $.

\subsection{Fully dressed function}
\label{s:full}

We now carry out the same Fourier transforms with the renormalized
function $\Gamma$, i. e. with
(\eqref{gcff} with $m=0$)
\beq
\label{gamx}
\Gamma(\xt) = \G_c(\xt|0) = \G^0(\xt|0)\  \fr[\al \log (|\xt| \teff)] \;,
\eeq
with the renormalization function
$\fr(l)$ given as a power expansion
(cf. \eqref{ffpow})
\beq
\fr(l) = \sum_{n=0}^{\infty} \afr_n \ l^n \;.
\eeq
 The Fourier transform of $\Gamma$ 
can be carried out as in \eqref{g0k2}, and the
 integral over $\phi'$ gives the same result, as $\fr$ only depends on
 the modulus of $\xt$. Thus, we are left with
\beq
\label{gamk}
\Gamma(\kom) = 
\frac{- i a^{\al}}{2 \pi}\ |\kom|^{\al-1}\ e^{-i \th}
\int_{|\kom| a}^{\infty} s^{-\al} d s \  
[- 2 \pi i J_1(s)]\  
\sum_{n=0}^{\infty}
 \afr_n \al^n \  \left(\log s + \log\frac{\teff}{|\kom|}\right)^n \;.
\eeq
Since $\fr$ is, in general, a complicated function, and its
coefficients $\afr_n$ very general, the Fourier
transform can be only carried up to the leading logarithmic behavior,
which, as discussed in Sec.~\ref{s:prob},
 amounts to considering 
 $l=\al \log (|\xt| \teff)$ of order $1$ but
$\al$ small.
 In this way, we can neglect the $\log s$ within braces in
\eqref{gamk} and  the effect of the renormalization function $\fr$
becomes merely multiplicative, provided one replaces $|\xt|$ with
$|\kom|^{-1}$ in its argument. We thus obtain
\beq
\label{gamk1}
\Gamma(\kom) = 
- \frac{|a\kom|^{\al}}{\kom\cdot \vv}\  
\fr\left(\al  \log\frac{\teff}{|\kom|}\right) 
= \G^0(\kom) \ \fr\left(\al  \log\frac{\teff}{|\kom|}\right) \;,
\eeq
where we have replaced the coefficient $g_{\al}$ with its $\al\to0$
limit  $g_{\al=0}=1$, consistently with the leading-logarithmic
approach. One can also verify {\it a posteriori} that inserting the
asymptotic result for $\fr(l)$ ($\sim e^l$) in \eqref{gamx} one indeed
obtains \eqref{gamk1} for small $\al$.

We now need  $\td(\xt,0)$ in real space, i. e. 
the
Fourier transform of $ - \Gamma(\kom)^{-1} $.
To express $ - \Gamma(\kom)^{-1} $ we
need
the reciprocal function of $\fr$ in terms of  its power-series
coefficients
$\abfr_n$:
\beq
\label{bfr}
\bfr(l) \equiv \frac{1}{\fr(l)} = \sum_{n=0}^{\infty} \abfr_n l^n \;,
\eeq
where  $\abfr_n$ can be determined from all the $\afr_m$ with
$m\leq n$.
Again, this function does not depend on angles, and we can proceed as
for \eqref{tdun1}, yielding
\beq
\label{tdx}
\td(\xt,0) \approx 
\frac{|\xt|^{\al-3}\ e^{i \phi}}{4 \pi^2\ g_{\al}\ a^{\al}}
\int_0^{\infty} s^{2-\al}\ d s \ 2 \pi i\ J_1(s)\  
\sum_{n=0}^{\infty}
 \abfr_n\ \al^n\  \left[\log(|\xt| \teff) - \log s \right]^n \;.
\eeq
The procedure is now slightly more complicated than for \eqref{gamk1},
since we have to consider terms at the first order in $\log s$.
The reason is that, if we  neglect completely 
the $\log s$ term, the
integral in \eqref{tdx} 
is  of order $\al$:
\beq
\int_0^{\infty} s^{2-\al} J_1(s)\ d s = g_{\al-2} \approx 2 \al \;.
\eeq
On the other hand, expanding the $[\cdots]^n$ power on the r.h.s. of \eqref{tdx},
and keeping
 the first term in $\log s$ yields a result of
order $1$:
\beq
\int_0^{\infty} s^{2-\al} \log s\  J_1(s)\  d s = - \frac{d}{d \al}
g_{\al-2} \approx -2  \;.
\eeq
The first integral,
 thus, gives a contribution 
$2 \ \abfr_n \  \al^{n+1} \ \left[\log(|\xt| \teff)\right]^n$
to the 
$n$-th term of the series
in \eqref{tdx},
 while the second gives
$2 \ n \ \abfr_n \ \al^{n} \ \left[\log(|\xt|
  \teff)\right]^{n-1}$. Both terms
 are of the same
order within the leading-logarithmic approach and must be taken into account.
We thus obtain
\beq
\label{tdx1}
\td(\xt,0) \approx 
\frac{i \al}{\pi\  a^{\al}}\ 
|\xt|^{\al-3}\  e^{i \phi}\  
\left[ 
\bfr(\al \log(|\xt|\teff)) + \bfr'(\al \log(|\xt|\teff))
\right]
 \;,
\eeq
since $\sum_{n=0}^{\infty} \abfr_n  \left( l^n+ n \ l^{n-1}\right) = 
\bfr(l)+\bfr'(l)$ (here, $\bfr'(l)=\frac{d}{d l} \bfr(l)$).
\eqref{tdx1}
 is the final result of this Section, which we need to insert in \eqref{ffm}.

\subsection{First $1/D$ corrections: 
irrelevance of $\perp$--nonlocal dressed hopping}
\label{s:nonl}

In the $D\to \infty$ limit, only the local effective hopping
$\td(\xt,\xy=0)$ 
is needed in the diagrams of Fig.~\ref{diagdinf}, as
$\perp$--nonlocal contribution vanish in this limit~\cite{dinf}.
In order to study the contribution of
finite--$D$ corrections, we consider 
the $\perp$--nonlocal contributions to \td, given by 
\beq
\label{tdnonl}
\td(\kom,\xy\not=0) = \int d \cp \ \D_{\xy}(\cp)\  \td(\kom,\cp) \;,
\eeq
where, since $\td$ depends on $\ky$ only through $\cp$,
we have introduced
 the ``generalized density of states'' (here, we use $R$ instead of $\xy$)
\beq
\label{dxy}
\D_{R}(\cp) = 
\int \left(\prod_{d=1}^{D'} \frac{d \ k_d}{2 \pi} \ 
e^{i    k_d R_d}\right)\  
\delta\left(\cp - \frac{2}{\resc} \sum_{d=1}^{D'} \cos k_d\right) \;.
\eeq
Following Refs.~\onlinecite{dinf}, we now introduce the Fourier
representation of the $\delta$ function,
 obtaining
\beq
\label{dxy1}
\D_{R}(\cp) = \int \frac{d \ s}{2 \pi} \ e^{i s \cp}\ 
I(s,R) \;,
\eeq
where the integral $I(s,R)$ is given by
\beq
\label{isr}
I(s,R)=
\int \prod_d \ \frac{d \ k_d}{2 \pi} \ 
e^{i    k_d R_d - 2 i s \cos k_d /\resc} 
\approx
\prod_d  \int \frac{d \ k}{2 \pi} \ e^{i k R_d} \ \left(1 - \frac{2 i s}{\sqrt{D'}} \cos k -
  \frac{2 s^2}{D'} \  \cos^2 k + O(D'^{-3/2})\right) \;,
\eeq
and we have expanded in powers of 
$1/\resc$.
The last integral gives at the leading order
\beq
\label{bigl}
\left\{ \matrix{ 1-\frac{s^2}{D'}         & \hbox{ for } R_d=0 \cr 
                 - \frac{i s}{\sqrt{D'}}  & \hbox{ for } R_d=1 \cr
                 - \frac{s^2}{2\ D'}      & \hbox{ for } R_d=2 \cr
               } \right. \;,
\eeq
and, in general, a term of order $(s/\resc)^n$ for $R_d=n$.
Inserting these results in \eqref{dxy1}, one obtains at the leading order in
$1/\resc$
\beq
\label{dxy2}
\D_{R}(\cp) = 
\int \frac{d \ s}{2 \pi} \ e^{i s \cp- s^2} \; 
 \prod_d a_{R_d}  \left(\frac{s}{\resc}\right)^{R_d} 
\;,
\eeq
where the $a_R$ are  coefficients obtained from \eqref{isr}. For
example, from \eqref{bigl}
 $a_0=1$, $a_1=-i$, $a_2=-1/2$.
The powers of $s$ in \eqref{dxy2} can be replaced with derivatives
with respect to $\cp$, yielding
\beq
\label{dxy3}
\D_{R}(\cp) = 
 \prod_d a_{R_d}\   \left( \frac{-i}{\resc} \ \frac{d}{d \ \cp}\right)^{R_d} 
\D(\cp) \;,
\eeq
where the usual density of states is given in \eqref{dens}.
We can now
insert \eqref{td}
 and \eqref{dxy3}, in \eqref{tdnonl}.
Since $\int d \cp \ \left(\frac{d}{d \ \cp}\right)^{n} \D(\cp) =0$ for 
$n\geq1$, for the nonlocal $\td$ we can subtract a $\cp$-independent term from
$\td(\kom,\cp)$, and write 
\beq
\label{tdnonl1}
\td(\kom,\xy\not=0) \propto 
\int d \cp \ 
\left[\td(\kom,\cp) + \frac{1}{\Gamma(\kom)}\right]
\left(\frac{1}{\resc} \ \frac{d}{ d \ \cp}\right)^{\overline{\xy}} \ 
\D(\cp) \;,
\eeq
where $\overline{\xy}\equiv \sum_{d} \xy_d$.
The  term within brackets in \eqref{tdnonl1}
 goes like $\Gamma(\kom)^{-2}$ 
for large $\Gamma(\kom)$, and the same holds for the integral 
 (the fact that the coefficient of $\Gamma(\kom)^{-2}$
diverges at $\cp=0$ might, at most, give a $\log$ correction).
Thus, $\td(\kom,\xy\not=0)$
vanishes at least like $\Gamma(\kom)^{-2}$
 for low
energies, i. e. faster than $\td(\kom,\xy=0)$.
Diagrams containing
$\perp$-nonlocal \td contributions are thus irrelevant in the
renormalization--group sense.

\section{Integration over center of mass coordinates}
\label{intcent}


In this section, we prove 
 \eqref{inty}, for the integration over the center of mass coordinate
 $\yt_{m+1}$.
In order to simplify the notation, we introduce the shorthand
$C(0,\cdots,{n}) \equiv \G^0_c(\yt_0', \cdots, \yt_{{n}}'
| \yt_{0},  \cdots, \yt_{{n}})$,
for the cumulants,
and
$G(0,\cdots,{n}) \equiv \G^0(\yt_0', \cdots, \yt_{{n}}'
| \yt_{0},  \cdots, \yt_{{n}})$ 
for the disconnected LL Green's function.
Notice that in these Green's functions
the implicit $\g$ and $r$ variables\cite{units} 
 are pairwise equal. More precisely, 
 $\g_k$ and $r_k$ , associated with
$\yt_k$, are equal to $\g_k'$ and $r_k'$,
 associated with $\yt_k'$.
The reason is that a $\td$ (or $\tp$) line does not change neither
$\g$ nor  $r$,
see Fig.~\ref{diagdinf}.
In addition, we define
$\xt_k = \yt_k'-\yt_k$, 
 $F({n})\equiv 2  \pi   |\xt_{{n}}|^2 \ G({n})$,
 $\quad \quad l_{j,{n}}\equiv 
\al \log\frac{|\xt_j|}{|\xt_{n}|} \left( \delta_{r_j,r_{n}} + \frac{1}{S}
  \delta_{r_j,-r_{n}} \right) $, and for the integration 
$\int\limits_{0\Reg  {n}} d^2 \yt_{{n}} $ we use the notation
$\int_{n}$.

The proof proceeds in two steps. We first show that the term on the
r.h.s. of \eqref{inty} is also given by an integral of disconnected Green's 
function, namely
\beqn
\label{intygf}
&& 
\int_{{n}} \left[ 
G(0,\cdots,{n}) -  G(0,\cdots,{n}-1)\  G({n}) \right]
\\ &&  \nonumber
=
G(0,\cdots,{n}-1)\  F({n}) 
\sum_{j=0}^{{n}-1} l_{j,{n}} \;,
\eeqn
where, for convenience, we have renamed $m\to {n}-1$.
Then, in Sec.~\ref{s:cumul},
 we prove the step from
\eqref{intygf} to
\eqref{inty} by induction.

\subsection{ Disconnected Green's function}
\label{s:dgf}

To show the first part,
we write 
the ${n}+1$-particle correlation function \eqref{g0ph} 
by using 
 \eqref{gll} in the following form
\beq
G(0,\cdots,{n}) =  G(0,\cdots,{n}-1) \ G({n}) \times 
\prod_{i=0}^{{n}-1} \frac
{P_{r_{i},r_{{n}}}(\yt_{i}-\yt_{n}')   P_{r_{i},r_{{n}}}(\yt_{i}'-\yt_{n}) }
{P_{r_{i},r_{{n}}}(\yt_{i}-\yt_{n})   P_{r_{i},r_{{n}}}(\yt_{i}'-\yt_{n}') }  \,
\eeq
where we have used the fact that $r_k=r_k'$. 
We thus have
\beq
\label{intm}
\int_{{n}}  \left[ 
G(0,\cdots,{n}) -  G(0,\cdots,{n}-1) \ G({n}) \right]
= G(0,\cdots,{n}-1) \ G({n}) \int_{n} I({n}-1) \;,
\eeq
with the argument of the integral
\beq
\label{Im1}
I({n}-1) \equiv 
\left[\prod_{i=0}^{{n}-1} \frac
{P_{r_{i},r_{{n}}}(\yt_{i}-\yt_{n}')   P_{r_{i},r_{{n}}}(\yt_{i}'-\yt_{n}) }
{P_{r_{i},r_{{n}}}(\yt_{i}-\yt_{n})   P_{r_{i},r_{{n}}}(\yt_{i}'-\yt_{n}') } -1\right] \;.
\eeq

The integral in \eqref{intm} is restricted to 
the region $0\Reg {n}$,
where $\xt_{n}$ is smaller than all other distances, which are the arguments of
the $P$'s, in
\eqref{Im1}. 
For this reason,
 we can expand $I({n}-1)$
 in powers of $\xt_{n}=\yt_{n}'-\yt_{n}$.
The zeroth order of this expansion 
is zero, as  $I({n}-1)=0$ for $\xt_{n}=0$.
The first order $I(n-1)^{(1)}$ gives
\beq
\label{Im11}
I({n}-1)^{(1)} = \sum_{i=0}^{{n}-1} \left[ -\xt_{n} \cdot 
\frac{\grad P_{r_i,r_{n}}(\yt_i-\yt_{n})}{P_{r_i,r_{n}}(\yt_i-\yt_{n})} 
+ \xt_{n} \cdot
\frac{\grad P_{r_i,r_{n}}(\yt_i'-\yt_{n})}{P_{r_i,r_{n}}(\yt_i'-\yt_{n})}
\right] \;,
\eeq
where the $\grad$ is considered as applied to the argument of the function.
Having in mind to integrate this expression
over $\yt_{n}$, one would be tempted to carry out a
shift in coordinates
 $\yt_{n} \to \yt_{n}+\xt_i$ in the second term within brackets in
\eqref{Im11}, thus obtaining zero.
This shift, however, has to be carried out with some care, since
 the logarithmic gradients
$\frac{\grad P_{r_i,r_{n}}(\yt_i-\yt_{n})}{P_{r_i,r_{n}}(\yt_i-\yt_{n})}$ 
 go like $1/|\yt_{n}|$ for large $|\yt_{n}|$. Therefore,
the integral
of each separate term in \eqref{Im11} does not converge, i. e. the
shift is not allowed without further prescriptions.
 However, this holds if one uses  the zero--temperature form \eqref{rxt0}.
On the other hand, the finite--temperature prescription \eqref{rxt}
 introduces a cutoff for values of {\it  each} of 
the arguments in the $\grad P/P$ in \eqref{Im11} of the order of
 $1/T$, 
making
the {\it separate} integrals absolutely convergent and allowing for
 the coordinate shift.
\taglia{ A simple way to see this is to replace \eqref{rxt} with 
a ``rotation symmetric one'' $\R(\xt) = \be \left(e^{|\xt|/be}
  -1\right)$, which has the correct properties: (i) exponential
divergence (so that $P$ decays exponentially)
for $|\xt|\gg \be$, and $R\propto |\xt|$ for $|\xt|\ll \be$.
Notice that it would be incorrect to take the zero--temperature 
form {\it before} carrying out the integral of $I({n}-1)^{(1)}$, since in this 
case, one would have neglected the fact that the temperature cutoff in the 
two terms in \eqref{Im11} occurs at different values of $\yt_{n}$.
}

We thus need to expand $I({n}-1)$ up to  the second order in $\xt_{n}$:
\beq
\label{Im12a}
1+I({n}-1) \approx 
\prod_{i=0}^{{n}-1} 
\left(1-\xt_{n}^{\aal} \ P_{\aal}(\bar \yt_i) +
    \xt_{n}^{\aal} \ \xt_{n}^{\bbe} \ P_{\aal,\bbe}(\bar \yt_i) \right)
\prod_{j=0}^{{n}-1} 
\left(1-\xt_{n}^{\aal'} \ P_{\aal'}(\bar \yt_j') +
    \xt_{n}^{\aal'} \ \xt_{n}^{\bbe'} \ P_{\aal',\bbe'}(\bar \yt_j')
 \right)^{-1} \;,
\eeq
where a sum over repeated indices $\aal,\aal',\bbe,\bbe'$ is understood, and
where
we have introduced the notations 
$\bar \yt_i \equiv \yt_i-\yt_{n}$, and $\bar \yt_i' \equiv
\yt_i'-\yt_{n}$.
Moreover, 
$\xt_{n}^{\aal}$ is the $\aal$ component of the vector $\xt_{n}$,  
$\quad P_{\aal}(\yt) = \frac{\de/\de \yt^\aal P(\yt)}{P(\yt)}$, and
$P_{\aal,\bbe}(\yt) = \frac{\de/\de\yt^\aal \de/\de\yt^{\bbe} P(\yt)}{2\
  P(\yt)}$. Moreover, we have omitted the $r_i$ indices in the functions
$P$, since they are fixed by
their $\yt$ arguments
[i. e., $P(\yt_i-\yt_{n})\equiv P_{r_i,r_{n}}(\yt_i-\yt_{n})$].

Expanding the denominator of 
 \eqref{Im12a} we obtain
\beqn
\label{Im12b}
1+I({n}-1) \approx && 
\prod_{i,j=0}^{{n}-1}
\left(1-\xt_{n}^{\aal} \ P_{\aal}(\bar \yt_i) +
    \xt_{n}^{\aal} \ \xt_{n}^{\bbe} \ P_{\aal,\bbe}(\bar \yt_i) \right)
\\ \nonumber && \times 
\left(1+\xt_{n}^{\aal'} \ P_{\aal'}(\bar \yt_j') -
   \xt_{n}^{\aal'} \ \xt_{n}^{\bbe'} \ P_{\aal',\bbe'}(\bar \yt_j')
+ \xt_{n}^{\aal'} \ \xt_{n}^{\bbe'} P_{\aal'}(\bar \yt_j') \ P_{\bbe'}(\bar \yt_j')
 \right) \;.
\eeqn
Collecting powers of $\xt_n^2$, we obtain 
the second--order term $I({n}-1)^{(2)}$:
\beqn
\nonumber
 && I({n}-1)^{(2)} = 
\xt_{n}^{\aal} \ \xt_{n}^{\bbe} \ \Bigl\{ \sum_i 
\left[ P_{\aal,\bbe}(\bar \yt_i) - P_{\aal,\bbe}(\bar \yt_i') +
   P_{\aal}(\bar \yt_i') \ P_{\bbe}(\bar \yt_i') \right]
\\ && \nonumber
+ \sum_{i>j} \left[ P_{\aal}(\bar \yt_i)  \ P_{\bbe}(\bar \yt_j)  +
           P_{\aal}(\bar \yt_i')  \ P_{\bbe}(\bar \yt_j') \right]
- \sum_{i,j} P_{\aal}(\bar \yt_i)  \ P_{\bbe}(\bar \yt_j') 
\Bigr\}
\\ && \label{Im12d}
= 
\xt_{n}^{\aal} \ \xt_{n}^{\bbe} \ \Bigl\{ \sum_i 
\left[ P_{\aal,\bbe}(\bar \yt_i) - P_{\aal,\bbe}(\bar \yt_i') +
  \left( P_{\aal}(\bar \yt_i') - P_{\aal}(\bar \yt_i) \right) 
                          \  P_{\bbe}(\bar \yt_i') \right]
\\ && \nonumber
+ \frac12 \sum_{i\not=j} \left[ P_{\aal}(\bar \yt_i)  \ P_{\bbe}(\bar \yt_j)  +
           P_{\aal}(\bar \yt_i')  \ P_{\bbe}(\bar \yt_j') 
  -   P_{\aal}(\bar \yt_i)  \ P_{\bbe}(\bar \yt_j') 
 -  P_{\aal}(\bar \yt_i')  \ P_{\bbe}(\bar \yt_j) 
\right]
\Bigr\} \;.
\eeqn
With the integration over $\yt_{n}$ in mind, and with the same arguments
about convergence as for 
\eqref{Im11}, we can carry out a coordinate  
shift of $\yt_{n}$ in some of the terms
of the sum \eqref{Im12d}.
First of all, we shift $\yt_{n}\to \yt_{n} + \xt_i$ 
in the 
$P_{\aal,\bbe}(\bar \yt_i')$ 
term, so that it becomes $P_{\aal,\bbe}(\bar \yt_i)$  and it cancels the
first $P_{\aal,\bbe}$ term.
Next,
 we transform the first-derivative term in the first sum 
in \eqref{Im12d} in the following way
\beqn
\nonumber
&&
 \left( P_{\aal}(\bar \yt_i') - P_{\aal}(\bar \yt_i) \right) 
                       \     P_{\bbe}(\bar \yt_i') 
\\ && \label{palbe1}
=  
\frac12 \ P_{\aal}(\bar \yt_i') \ P_{\bbe}(\bar \yt_i') 
+ \frac12 \ P_{\aal}(\bar \yt_i')  \ P_{\bbe}(\bar \yt_i') 
- \frac12 \ P_{\aal}(\bar \yt_i) \ P_{\bbe}(\bar \yt_i') 
- \frac12 \ P_{\aal}(\bar \yt_i) \ P_{\bbe}(\bar \yt_i') 
\\  && \label{palbe2}
\to 
\frac12  \ P_{\aal}(\bar \yt_i') \ P_{\bbe}(\bar \yt_i') +
\frac12 \ P_{\aal}(\bar \yt_i) \ P_{\bbe}(\bar \yt_i) 
- \frac12 \ P_{\aal}(\bar \yt_i) \ P_{\bbe}(\bar \yt_i') 
- \frac12 \ P_{\aal}(\bar \yt_i') \ P_{\bbe}(\bar \yt_i)  
\\ \nonumber &&
= 
\frac12 \ \left[ P_{\aal}(\bar \yt_i) - P_{\aal}(\bar \yt_i') \right]  
       \  \left[ P_{\bbe}(\bar \yt_i) - P_{\bbe}(\bar \yt_i') \right]
         \;,
\eeqn
where \eqref{palbe2} is obtained 
by shifting $\yt_{n} \to \yt_{n} + \xt_i$ in the second term 
 and by exchanging $\aal$ and $\bbe$ in the 
fourth term of \eqref{palbe1} (which is allowed, as \eqref{Im12d} is
  symmetric in $\aal,\bbe$).
Inserting the result in \eqref{Im12d}, and factorizing in the same way 
the terms in
the last sum, we finally get
\beqn
\label{Im12final}
  I({n}-1)^{(2)} \to &&
\frac12 \ \xt_{n}^{\aal}\  \xt_{n}^{\bbe}\  \sum_{i,j}
\left[ P_{\aal}(\bar \yt_i) - P_{\aal}(\bar \yt_i') \right]  
      \    \left[ P_{\bbe}(\bar \yt_j) - P_{\bbe}(\bar \yt_j') \right] 
      \\ \nonumber &&
= \frac12 \left\{ \sum_{\aal,i} \xt_{n}^{\aal}\ 
\left[P_{\aal}(\bar \yt_i) - P_{\aal}(\bar \yt_i') \right]
\right\}^2 \;,
\eeqn
where ``$\to$'' means that it is equal but for a shift of the
integration variable $\yt_n$ in some of the summands.

We now need the logarithmic gradients $P_{\aal}(\yt_a-\yt_b)$.
If the points $a$ and $b$ correspond to two electrons on opposite side
of the FS, i. e., $r_a=-r_b$, then from
Eqs.~(\ref{P},\ref{pexp},\ref{rxt0}), 
$P^0(\yt) = |\yt|^{-B}$, 
and
\beq
P^0_{\aal}(\yt) = \frac{\de_{\aal} |\yt|^{-B} }{|\yt|^{-B}}
= - B \frac{\yt^{\aal}}{|\yt|^2} \;,
\eeq
where we have 
 set $\yt=\yt_a-\yt_b$, and
introduced a superscript symbol $0$ or $1$ to $P_{\aal}$,
depending on whether  $r_a=-r_b$  or $r_a=r_b$, respectively.
In the second case,
 $r_a=r_b\equiv r$, we can write
$P^1(\yt) = c \ |\yt|^{-2-\al} \ \yt \cdot \vv $, where
$c$ is a constant, and the two-component vector $\vv = (1,-i \ r)$,
slightly different from  the one defined in Sec.~\ref{s:corr}.
Differentiating, we obtain
\beq
P^1_{\aal}(\yt) = -(2+\al) \frac{\yt^{\aal}}{|\yt|^2} + \frac{\vv^{\aal}}{\yt
  \cdot \vv} =
\frac{1}{|\yt|^2} \left[ (\yt \cdot \vv^*) \vv^{\aal} - (2+\al)
  \yt^{\aal} \right]\;,
\eeq
as $|\yt|^2 = (\yt \cdot \vv ) \ (\yt \cdot \vv^* )$. 

There are thus three types of integrals to be carried out in
\eqref{intm} with the second--order term \eqref{Im12final}.
First, for the case that $r_{n}=-r_{i}=-r_{j}$, we need
an integral of the form
\beq
\label{intrarb1}
\int_{n}   P^0_{\aal}(\yt_{i}-\yt_{n})\  P^0_{\bbe}(\yt_{j}-\yt_{n})
= B^2 \int d^2 \yt\  \frac{\yt^{\aal} - \xt^{\aal}}{|\yt - \xt|^2}
  \frac{  \yt^{\bbe}}{|\yt|^2} = 
B^2 \pi\  \delta_{\aal\bbe} 
\log\frac{R}{\max(|\yt_{i}-\yt_{j}|,|\xt_{n}|)}
\;,
\eeq
where in the intermediate step 
we have transformed $\yt_{n} = \yt + \yt_{j}$, and
$\yt_{i}-\yt_{j}=\xt$, 
and used the result \eqref{logint2}.
Here, we have introduced a large--distance cutoff $R\propto 1/T$,
 from which, eventually, the
final result does not depend. 
\taglia{
Since the partial expressions only
depend logarithmically on $R$, this is equivalent to taking 
the temperature cutoff for each of the $P$, as discussed below
\eqref{Im11}.
\check{check?}
}
The max in the logarithm practically only applies when $i=j$, as
$|\xt_m|$ is always the smallest distance. In this case, the result is 
obtained by
 keeping in mind that $|\xt_{n}|$ is the
short-distance cutoff, and by applying \eqref{logint1}.
For  $r_{n}=r_{i}=r_{j}=r$, we need
\beqn
\label{intrarb2}
&&
\int_{n} P^1_{\aal}(\yt_{i}-\yt_{n})\ P^1_{\bbe}(\yt_{j}-\yt_{n})
\\ \nonumber &&
=
\int_{n} 
\frac{1}{|\yt|^2 |\yt-\xt|^2}
\left\{\left[(\yt-\xt)\cdot \vv^*\right]\  \vv^{\aal} - (2+\al)\ 
  (\yt^{\aal}-\xt^{\aal})\right\} \ 
\left[ (\yt\cdot \vv^*) \ \vv^{\bbe} - (2+\al)\ 
  \yt^{\bbe}\right]  = 
\\ \nonumber &&
\left[ \vv^{*\aal'} \ \vv^{\aal} - (2+\al)\  \del_{\aal\aal'}\right]\ 
\left[ \vv^{*\bbe'} \ \vv^{\bbe} - (2+\al)\  \del_{\bbe\bbe'} \right]
\int d^2 \yt\  \frac{\yt^{\aal'} - \xt^{\aal'}}{|\yt - \xt|^2}
  \frac{  \yt^{\bbe'}}{|\yt|^2}
\\ \nonumber &&
=
\pi \  (2+\al)\ \al \  \del_{\aal\bbe} \ 
\log\frac{R}{\max(|\yt_{i}-\yt_{j}|,|\xt_{n}|)}
 \;,
\eeqn
again using \eqref{logint2} and the fact that
 $ \vv^{*\aal'} \ \vv^{*\aal'}=0$, and
$\vv^{\aal} \ \vv^{*\bbe}+ \vv^{*\aal}\  \vv^{\bbe} = 2 \del_{\aal\bbe} $.
Finally, for
$r_{n}=r_{j}=-r_{i}$, we have
\beqn
\label{intrarb3}
&&
\int_{n}  P^0_{\aal}(\yt_{i}-\yt_{n})\  P^1_{\bbe}(\yt_{j}-\yt_{n})
=
( -B)\ 
\left[ \vv^{*\bbe'} \vv^{\bbe} - (2+\al)\  \del_{\bbe'\bbe}\right]\ 
\pi\ \del_{\aal,\bbe'}\  \log\frac{R}{\max(|\yt_{i}-\yt_{j}|,|\xt_{n}|)} \;.
\\ \nonumber &&
=
\pi \left[ B (2+\al)\   \del_{\aal\bbe} - B\  \vv^{*\aal}\ \vv^{\bbe}\right]
\ \log\frac{R}{\max(|\yt_{i}-\yt_{j}|,|\xt_{n}|)} 
\to
\pi\  (1+\al) \ B\ \del_{\aal\bbe} \ 
\log\frac{R}{\max(|\yt_{i}-\yt_{j}|,|\xt_{n}|)} 
\;,
\eeqn
where, in the last step we have symmetrized with respect to $\aal$ and $\bbe$.

We can thus use these results to integrate \eqref{Im12final}
and obtain
\beqn
\nonumber
\int_{n}  I({n}-1)^{(2)}  \approx &&
\frac{\pi}2\ \del_{\aal\bbe}\ \xt_{n}^{\aal} \ \xt_{n}^{\bbe} \Bigl\{
\sum_{i\not=j=0}^{{n}-1}
  \log \frac{|\yt_i' - \yt_j| |\yt_i - \yt_j'|}{|\yt_i - \yt_j|
    |\yt_i' - \yt_j'|} \left[ \left( (2+\al)\ \al \ \delta_{r_i,r_{n}} + B^2 \
  \delta_{r_i,-r_{n}} \right) \ \delta_{r_i,r_j} +  (1+\al) \ B\ \delta_{r_i,-r_j}
  \right]
\\ \label{intIm}
&& 
+ \sum_{i=0}^{{n}-1}
 2 \ \left( 
 (2+\al)\ \al \  \delta_{r_i,r_{n}} 
+
B^2 \delta_{r_i,-r_{n}} 
\right)
\  \log \frac{|\xt_i|}{|\xt_{n}|}
\Bigr\}
\;,
\eeqn
where we have considered the case $i=j$ separately, and used the fact
that $|\xt_{n}|$ is (much~\cite{great})
 smaller than all other distances in
 the region
${0\Reg {n}}$, and, thus, it can be neglected whenever
 it appears summed to other distances 
as the argument of a logarithm.
Notice that the large-distance cutoff $R$ cancels out, as anticipated.

Consider now the terms in \eqref{intIm} with $i\not=j$. These give
logarithmic contributions of the form
\beq
\label{logij}
  \log \frac{|\yt_i + \xt_i - \yt_j| |\yt_i - \yt_j -\xt_j|}
            {|\yt_i - \yt_j|         |\yt_i + \xt_i - \yt_j -\xt_j|}
\eeq
 For the sake of definiteness,
  let's take $i>j$ in \eqref{logij}, so that, in the relevant region
  $0\Reg {n}$, $\xt_i$ is smaller than all other
  differences in the arguments of the logarithm and thus can be set to 
  zero. In this way, numerator and denominator in \eqref{logij}
            cancel and the result is zero.
This means that the terms with $i\not=j$ in \eqref{intIm} do not
            contribute to the  leading logarithmic divergence.
Thus, the only contribution to \eqref{intIm} stems from the second
term within brackets, which gives
\beq
\label{intIfinal}
\int_{n} I({n}-1)^{(2)} \approx 
2  \pi |\xt_{n}|^2 \sum_{i=0}^{{n}-1} 
 \al \log \frac{|\xt_i|}{|\xt_{n}|} 
 \left( \delta_{r_i,r_{n}} + \frac{1}{S}
  \delta_{r_i,-r_{n}} \right)
\;,
\eeq
where we have taken
 $B^2\approx 2 \al/S$, and $(2 + \al) \ \al \approx 2 \al$,
consistently with 
the leading-log  approximation.
Inserting \eqref{intIfinal} into \eqref{intm} yields the desired
result \eqref{intygf}.
$\cdots$ well, there is certainly a much faster and elegant 
way to get the rather
simple result \eqref{intIfinal}!.

\subsection{Some logarithmic integrals}
\label{logint}

Here, we evaluate the integrals  used in \eqref{intrarb1}, and following.
The first integral is straightforward
\beq
\label{logint1}
\int_{\Delta< |\yt| <R} d^2 \yt \ \frac{\yt^{\al}  \yt^{\be}}{|\yt|^4}
=  \pi \ \del_{\al\be} \ \log \frac{R}{\Delta}\;,
\eeq
where $R$ is a large-distance  and
$\Delta$ a short-distance cutoff
for $|\yt|$, which are needed
due to the logarithmic  divergences of the integral.
We next prove that
\beq
\label{logint2}
\int_{|\yt|<R} d^2  \yt \ \frac{\yt^{\al} - \xt^{\al}}{|\yt - \xt|^2} \ 
  \frac{  \yt^{\be}}{|\yt|^2} \approx  \pi \ \del_{\al\be} \ \log
  \frac{R}{|\xt|} \;,
\eeq
where $\approx$ means at the leading order in $\log (R/|\xt|)$.
The integral converges at short distances, so there is no need for
a short-distance cutoff, and diverges logarithmically at large distances.
We can split the integral in two regions:
(i) $ |\xt|  N \leq |\yt| \leq R$, and 
(ii) $  |\yt| \leq |\xt| N$, with $N$ large but much smaller than 
$R/|\xt|$,
 so that $\log N$ can be neglected.
In region (i), the integrand can be safely approximated by
$ \frac{\yt^{\al} \yt^{\be}}{|\yt|^4}$, whose integral, taken from
  \eqref{logint1}, gives 
$\pi\ \del_{\al\be} \log \frac{R}{|\xt|} $.
In region (ii), the only length scale left is $|\xt|$, since the
integral converges at short distances, and thus there is no
logarithmic contribution from this region, and
\eqref{logint2} is proven.

\subsection{Integration of cumulants}
\label{s:cumul}

We have thus proven \eqref{intygf}, 
an equation similar to 
\eqref{inty}, but with
 {\it disconnected} Green's functions instead of cumulants.
We now prove  by induction the same thing with cumulants.
Induction is the best way to do it, as
cumulants themselves can be  written
by induction
in terms 
of disconnected Green's functions. 
A $n$--particle
cumulant consist of the sum of  the $n$--particle disconnected Green's
function plus 
an appropriate sum of products of  $k$-particle Green's functions 
with $k<n$ (Ref.~\onlinecite{cumul}). 
However, we can show that in our problem, we can restrict to the so called 
``paired'' contributions to the cumulants, i. e. 
we can throw away all 
those terms in the sum 
 in which, 
for any $k$, the coordinates $\yt_k$ and $\yt_k+\xt_k$ do not belong to 
the same Green's function. The fact that these terms (``unpaired
terms'', see Fig.~\ref{pair}) can be neglected
is shown in Sec.~\ref{s:pair}.  

Let us  write  in the shorthand form
\beq
\label{intym1}
\int_{n}
C(0,\cdots,n) 
\approx
C(0,\cdots,n-1) \ F(n) \  
\sum_{j=0}^{n-1} l_{j,{n}}\;,
\eeq
which coincides with \eqref{inty} for 
$n=m+1$.
The induction procedure consists in  proving (i) that \eqref{intym1} holds for
$n=1$, and (ii)
that in the hypothesis that \eqref{intym1} holds for all $n\leq m$, it 
also holds for
$n=m+1$.

For $n=1$, $C(0,1)$ is equal to $G(0,1)-G(0)G(1)$ plus
 unpaired terms. Since, as discussed above,
unpaired terms can be neglected, for $n=1$
\eqref{intym1} coincides with
  \eqref{intygf}, which we have just shown in Sec.~\ref{s:dgf}.

We now assume \eqref{intym1} to be valid for all $n\leq m$.
Let us first
introduce the definition of a cumulant in terms of connected Green's
functions, 
\beq
\label{cumdef}
C(0,\cdots,n) = G(0,\cdots,n) - \sum_{P(0,\cdots,n)} 
C(P_1)  \cdots  C(P_{N_P}) \;,
\eeq
where we have already left out unpaired terms.
In \eqref{cumdef},
the $P_k$ are subsets of the set of integers $\{0,\cdots,n\}$.
$\quad \{P_1,\cdots,P_{N_P}\}$ 
is  a partition with $N_P$ terms of this set, 
and the sum 
($\displaystyle \sum_{P(0,\cdots,n)} $)
goes over
all inequivalent 
partitions with $N_P\geq2$ of this set.
Equivalent partitions are the ones which can be set equal by a permutation.
Introducing
\eqref{cumdef} in the result for the disconnected Green's functions
\eqref{intygf} (with $n\to m+1$),
one obtains
\beqn
\label{intpar}
&& \int_{m+1}
\left[ C(0,\cdots,m+1) 
- G(0,\cdots,m) \ G(m+1) + \sum_{P(0,\cdots,m+1)} 
  C(P_1)  \cdots  C(P_{N_P}) \right]
\\ &&  \nonumber 
= G(0,\cdots,m) \ F(m+1) \sum_{j=0}^m l_{j,m+1} \;.
\eeqn
In \eqref{intpar}, 
the sum over the partitions of the set of integers  $0,\cdots,m+1$ 
can be further splitted in the following way
\beqn
\nonumber
 \sum_{P(0,\cdots,m+1)}   \ C(P_1)  \cdots  C(P_{N_P})
= &&
\sum_{P(0,\cdots,m)}  \ 
\sum_{k=1}^{N_P} C(P_1) \cdots C(P_k,m+1) \cdots  C(P_{N_P})
\\ && \label{splitpart}
+ \sum_{P(0,\cdots,m)}  
 C(P_1) \cdots  C(P_{N_P}) \ C(m+1) 
+ C(0,\cdots,m) \ C(m+1) \;,
\eeqn
i. e., 
into  the sum over the partitions of the
integers  $0,\cdots,m$ with the element 
$m+1$ either appended in all subsets of the partition,
or taken alone.
Upon applying the definition \eqref{cumdef} with $n=m$ to the last
term 
 on the r.h.s.,   \eqref{splitpart} becomes
\beq
\label{splitpart1}
 \sum_{P(0,\cdots,m+1)}   \ C(P_1)  \cdots  C(P_{N_P})
= 
\sum_{P(0,\cdots,m)}  
\sum_{k=1}^{N_P} C(P_1) \cdots C(P_k,m+1) \cdots  C(P_{N_P})
+ G(0,\cdots,m) \ C(m+1) \;,
\eeq
which,  inserted into the l.h.s. of  \eqref{intpar},
cancels the second  term within brackets, giving
\beqn
\label{intpar1}
&&
 \int_{m+1}
 C(0,\cdots,m+1) = - \int_{m+1}
  \sum_{P(0,\cdots,m)} 
\sum_{k=1}^{N_P} C(P_1) \cdots C(P_k,m+1) \cdots  C(P_{N_P})
\\ \nonumber &&
+
 G(0,\cdots,m) \ F(m+1) \sum_{j=0}^m l_{j,m+1} \;.
\eeqn
The 
 integral $\int_{m+1}$ in the first term on the
r.h.s. of \eqref{intpar1}
can be evaluated by using
 the induction hypothesis \eqref{intym1} with $n\leq m$, 
as $C(P_k,m+1)$ is a cumulant with less than
$m+1$ particles.
We thus obtain for this term
\beqn
\label{intpar2}
&&
 - 
  \sum_{P(0,\cdots,m)} 
\sum_{k=1}^{N_P} C(P_1) \cdots \left[C(P_k) \ F(m+1) \sum_{j \in P_k}
l_{j,m+1} \right] \cdots  C(P_{N_P}) =
\\ \nonumber &&
 - \sum_{P(0,\cdots,m)} 
 C(P_1) \cdots C(P_k) 
 \cdots  C(P_{N_P}) 
\ F(m+1) \sum_{j=0}^m l_{j,m+1}  \;,
\eeqn
since 
$\sum_{k=1}^{N_P} \sum_{j \in P_k} = \sum_{j=0}^m$.
Inserting the last result
in \eqref{intpar1}, and using again the definition
 \eqref{cumdef} yields the desired result, i. e., \eqref{intym1} with
 $n=m+1$.

\subsection{Irrelevance of ``unpaired'' terms}
\label{s:pair}

We want to show that the ``unpaired terms'' in a cumulant do not
contribute to the leading logarithmic divergences in any of the terms of the
sum \eqref{gamma}.
A $m+1$-particle  cumulant is the sum of products of $n$-particles Green's
functions with $n\leq m+1$. For ``unpaired terms'' we mean those
terms in the sum  for which 
 some paired variables (i. e. $\yt_k$ and $\yt_k' \equiv\yt_k+\xt_k$)
do not belong to the same Green's function. For example, for $m=1$
\beq
G^0_c(0,1|0',1') =  G^0(0,1|0',1') - G^0(0|0')  \ G^0(1|1') +
G^0(0|1') \ 
G^0(1|0')  \;,
\eeq
the last term on the r.h.s. is unpaired, while the first two are
paired. The first two terms, when inserted in \eqref{gamma} give a
$\log^2(|\xt_0|\teff)$ contribution, as discussed in 
 Sec.~\ref{s:dinf}.
 The contribution to $\Gamma$ of the last term can be best
understood diagrammatically (see Fig.~\ref{pair}). Its
contribution, written in momentum space, is proportional to
\beq
\int d^2 \kom \ G^0(\kom) \ \td(\kom,\xy=0) \ G^0(\kom) \ e^{i \kom \cdot \xt_0}
\approx
- \int d^2 \kom \ \frac{G^0(\kom)}{\fr(\al \log \frac{\teff}{|\kom|})}
\  e^{i \kom \cdot \xt_0}
\approx \frac{G^0(\xt_0)}{\fr[\al \log (\teff |\xt_0|)]} \;,
\eeq
i. e., it does not give additional logarithmic terms to $\Gamma$, contrary
to the contribution from the paired terms.
The same thing happens at higher order, namely, while the 
paired terms
of  a $m+1$-particle cumulant inserted in \eqref{gamma} give a correction
 of order $\log^{2 m}(|\xt_0|\teff)$ to $\Gamma$, the unpaired terms give smaller
 powers of the logarithm and can thus be neglected at the leading
 logarithmic order.

\levfig{
\begin{figure}
   \centerline{\psfig{file=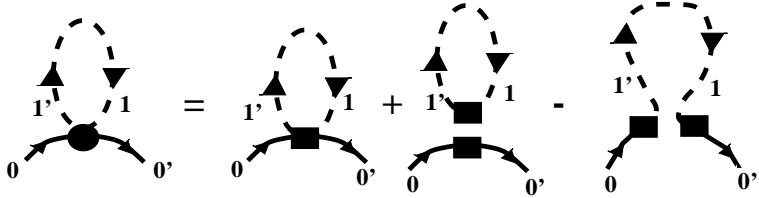,width=\figwidth}}
\caption{
\label{pair}
Splitting of a cumulant contribution into ``paired'' and ``unpaired''
terms.
A cumulant is indicated by the black dot and is obtained as a sum of
disconnected Green's functions, represented by black squares. The
last diagram on the r.h.s. 
is an ``unpaired'' one, according to the definition of
Sec.~\ref{s:pair}, and does not contribute to the leading logarithmic
divergences, as shown in that section. 
}
 \end{figure}
}

\section{Relevant integration region}
\label{proofrelevant}

In this section, we show that 
the leading contribution in $\log(|\xt_0| \teff)$ to each of  
the integrals in the series  \eqref{gamma}, restricted to
$|\xt_m| \leq |\xt_{m-1}| \leq \cdots \leq |\xt_1|$, can be further
restricted (A) to the subregion
$|\xt_1| \leq |\xt_0|$, and  (B) to 
$|\xt_p| < |\yt_q+\epsilon_1 \xt_q -\yt_{q'} - \epsilon_2 \xt_{q'}|$ 
for each $p \geq q,q' $, $q\not=q'$, and $\epsilon_i=0,1$.
This relevant region, which we  call 
``$0\Reg m$'', is the one  where the modulus of the relative
coordinate
$\xt_p$
 with
a given index $p$ ($p=0,\cdots,m$)
 is smaller than the distance between any two different points with
indices $q, q'$ 
smaller or equal than $p$.
The remaining regions do not contribute to the leading logarithmic
divergence of the integral. This is a crucial point in giving the
simple expression \eqref{inty}.

Let us start with $m=1$. We have already seen 
in Sec.~\ref{s:dinf}
that the contribution to 
\eqref{gamma} from the 
region $0\Reg 1$ gives a $\log^2$ term, and, for general $m$ one has a
$\log^{2m}$ contribution.
Consider now the integration region 
$|\xt_0| < |\xt_1|$
 in the term $m=1$ in \eqref{gamma}, which violates (A).
The  integration over the ``center of mass'' coordinate $\yt_1$
can be replaced with an integration over $\yt_0$, since the integrand
depends on the difference between  the two. As a consequence, one can just take
over the result \eqref{intygf} with $n=1$, and interchange
the labels $1$ and $0$.
This is correct because now $|\xt_0|$ is smaller than $|\xt_1|$.
One thus obtains
\beq
\int_1
  \left[ 
G(0,1) -  G(0) \ G(1) \right]
\propto 2 \ \al \  \pi \ G(0)\  G(1)\  |\xt_0|^2 \
\log\frac{|\xt_1|}{|\xt_0|}
 \;.
\eeq
If one now inserts the expression \eqref{tdx0} for the {\it \LO} $\td$, and
integrates
$\xt_1$ from $|\xt_1|=0$  to $|\xt_1|=|\xt_0|$, the result is
proportional to
$\al^2 \ G(0) \ |\xt_0|^2 \ \int_{|\xt_1|<|\xt_0|} \frac{d^2
  \xt_1}{\xt_1^4}
 \ \log
\frac{|\xt_1|}{|\xt_0|} = \al^2 G(0) \times O(1)$, where $O(1)$ is a
term of order unity, i. e. without
logarithmic contribution. For a generic term of the $\log$ expansion,
\eqref{tdx1},
of the {\it completely dressed} $\td$ one has a similar result, namely  after
integration over $\xt_1$ one has no additional $\log$ contribution, while
one gets a term $\al^2$, i. e. one ``looses'' two logarithms from
integrating in that region.

Taking now $m>1$,
one 
first integrates the variables $\yt_2,\xt_2,\cdots,\yt_m,\xt_m$ by
using \eqref{gc}. However, this integration simply renormalizes the
two-particle cumulant $\G_c$ by a factor of
order $1$ in the leading $\log$, i. e. by
 a sum of powers  of 
$(\al \log |\xt| \teff)$.
Now one can proceed integrating over $\yt_1, \xt_1$.
By the same argument as above, 
it is straightforward to show that
 integration from the region $|\xt_0|<|\xt_1|$ does not
contribute additional logarithms, while it gives  a term $\al^2$, and
can thus be neglected.
 We have thus proven (A), i. e. that the leading divergent
contribution to each term of the expansion \eqref{gamma} comes from
the region $|\xt_0|>|\xt_1|$.

To show the second part (B) of the statement,
we should first understand how a logarithmic contribution to
\eqref{gamma} comes
out. Let us  consider
the integration over $\yt_m$ of Sec.~\ref{s:dgf}.
 $I(m-1)$ (\eqref{Im1}), which is the only $\yt_m$-dependent part of
$G(0,\cdots,m)$,
 can be written in  the generic form
\beq
I(m-1)= \prod_{i} \PP_i(\yt_m-\rt_i) \;,
\eeq
where the $\PP_i$ are functions with an integrable  singularity
in $\v 0$,
[whenever the exponents $1+\al$, and $B$ (\eqref{defB}) 
are smaller than $2$].
Since the singularities are integrable, there is no divergent (power
law or logarithmic)
contribution from integration of $\yt_m$ in the neighborhood of the
points $\rt_i$. For the sake of definiteness,
let us suppose that 
$\rt_1$ and $\rt_2$ are the two nearest points among the $\rt_i$, and call
$\Delta\equiv|\rt_2-\rt_1|$ their distance.
Then, one can consider the circle
$R_{<\Delta}$ 
of radius $N \Delta$ around $\rt_1$, with $N$ some  number of the order $1$ 
smaller than the relevant logarithmic scale.
This region contains $\rt_1$ and $\rt_2$ but none of the  other
$\rt_i$ points
(In case 
 there are other points $\rt_i$ inside this region, 
the following argument does not change,
provided their
distance from $\rt_1$ and $\rt_2$ is neither much larger nor much
smaller than $\Delta$).
Inside $R_{<\Delta}$, there is just one characteristic
 length scale $\Delta$, since there is no
need for a short-distance scale  due to the convergence of the
integral. Thus, by simple dimensional analysis, one obtains for the
integration in this region
$\int_{R_{<\Delta}} d^2 \yt_m \ I(m-1) \propto \Delta^2$ with no $\log$
contribution
 since
one needs two length scales for a $\log$.
A logarithmic contribution can only come from integrating $\yt_m$ in the
 remaining region $R_{>\Delta}$, where more energy scales are
 available.
In this region, one can expand in powers of $\Delta$, when it appears
 as an argument of the $\PP_i$, as we have done in Sec.~\ref{s:dgf}
 whith $\Delta=|\xt_m|$.

Let us start from the simplest case
$m=1$. Here, there are four points $r_i$, namely
   $\yt_0$, $\yt_0-\xt_1$, $\yt_0'\equiv\yt_0+\xt_0-\xt_1$,
and $\yt_0+\xt_0$. 
Since we have~\cite{great}
$|\xt_1|\ll
|\xt_0|$,  the smallest distance $\Delta$ between two of these points is given by
$|\xt_1|$.
 As discussed above, the leading logarithmic contribution
to \eqref{gamma}
comes from the region $R_{>|\xt_1|}$, where
$|\yt_1-\yt_0|, |\yt_1-\yt_0'|,|\yt_1'-\yt_0|, |\yt_1'-\yt_0'| >
|\xt_1|$ 
which proves  result (B) for $m=1$. 
It is now straightforward to extend this argument by induction for any
$m$.
Specifically, we first assume that we can 
 restricts to the region where the distances
$|\yt_p-\yt_q|, |\yt_p-\yt_q'|,|\yt_p'-\yt_q'|$ (let's call them
``$|\yt_p-\yt_q|$ and primed'')
are larger than
$|\xt_{m-1}|$, for
$p,q\leq m-1$.
Then, since $|\xt_{m-1}|>|\xt_m|$ (we are restricting to the region
$1\reg m$), 
it remains to be shown that 
the region where any one of the distances
$|\yt_m-\yt_q|$ and primed
is smaller than
$|\xt_m|$ 
does not
contribute to the leading logarithmic divergence. 
Since $|\xt_m|$ is smaller than all {\it other} distances 
$|\yt_p-\yt_q|$ and primed, 
we can apply the argument above, according to which
 logarithmic contributions from the integral in $d^2 \yt_m$ 
come from the region outside of circles of radius $|\xt_m|$
 from any of the points $\yt_p$ or $\yt_p'$. This  proves the
 statement.

\section{Solution of the recursive equation by power expansion}
\label{s:solu}

In this section, we describe
 the practical procedure to solve the recursive set of equations 
\eqref{ff} by power expansion up to very high order. We also show some 
results of the corresponding Pad\'e resummation.
We expand the  functions $\FF_m$ in power of their arguments
\beq
\label{ffpow}
\FF_m(l,l_m) = \sum_{i,j=0}^{\infty} f^{(m)}_{i,j} \ l^i \ l_m^j \;.
\eeq
We can use two known results, namely (i)
$\FF_m(l,0)=1$, which implies
 $f^{(m)}_{i,0}=\delta_{i,0}$, and (ii)
$f^{(0)}_{i,j}=\delta_{i,0} \ \afr_j$, as 
$\FF_0(l,l_0)$ only depends on $l_0$.

Inserting \eqref{ffpow}, 
and the expansion for the reciprocal function \eqref{bfr}
in \eqref{ff} yields
\beqn
\label{ffpow1}
&&
\sum_{i,j=0}^{\infty} f^{(m)}_{i,j} \ l^i\  l_m^j =
  1+ 2(S+1) \  \int_0^{l_m} d l'\ 
 [l+l_m- (m+1) \ l']\ 
 \\ && \nonumber \times
\sum_{r,s=0}^{\infty} f^{(m+1)}_{r,s} \ (l+l_m)^r \ l'^s
\sum_{p=0}^{\infty} [\abfr_p + (p+1) \ \abfr_{p+1}] \ l'^p
 \\ && \nonumber =
  1+ 2(S+1)  \ \int_0^{l_m} d l'\ 
\sum_{r,s,p=0}^{\infty} f^{(m+1)}_{r,s} \ [\abfr_p + (p+1) \
\abfr_{p+1}] \   
\left[(l+l_m)^{r+1} \ l'^{s+p} - (m+1) \ (l+l_m)^{r} \ l'^{s+p+1} \right] \;.
\eeqn
Carrying out the integration and applying the binomial expansion
(with the agreement that ${r\choose s}=0$ for $s>r$),
 \eqref{ffpow1} becomes
\beqn
\label{ffpow2}
&&
 = 1+ 2(S+1)  \ 
\sum_{r,s,p=0}^{\infty} \ f^{(m+1)}_{r,s} \ [\abfr_p + (p+1) \ \abfr_{p+1}]  
\sum_{q=0}^{r+1} l^q \ l_m^{r+s+p+2-q} \  
\left[ {r+1\choose q} \  \frac{1}{s+p+1} - {r \choose q} \frac{m+1}{s+p+2}
 \right]
 \\ && \nonumber =
1 +  \sum_{i=0}^{\infty} \ \sum_{j=\max(1,2-i)}^{\infty} \ l^i \ l_m^j \ 
 B^{(m+1)}_{i,j} \;,
\eeqn
\taglia{ remember $i\leq r+1$ (due to the binomials) }
where we have 
replaced $q=i$ and $r+s+p+2-q=j$, and
introduced the coefficients
\beq
\label{ffpow3}
B^{(m+1)}_{i,j} = 2(S+1) \ \sum_{r=\max(i-1,0)}^{i+j-2} \ \sum_{p=0}^{i+j-r-2} 
\    f^{(m+1)}_{r,i+j-r-p-2} 
\ [\abfr_p + (p+1) \ \abfr_{p+1}]  \ 
\left[ {r+1\choose i} \  \frac{1}{i+j-r-1} - {r \choose i}
   \frac{m+1}{i+j-r} \right] \;.
\eeq
Comparison of \eqref{ffpow1} with \eqref{ffpow2} gives the relation
   between the $f^{(m)}$ and the $f^{(m+1)}$, namely
\beq
f^{(m)}_{i,j} = \cases{ B^{(m+1)}_{i,j},  & {\rm for} $i+j \geq 2,i\geq 0,j\geq1$ ; \cr
                1,  & {\rm for} $i=j=0$                     ; \cr 
                0,  & {\rm otherwise}                       ; \cr
}
\eeq

From \eqref{ffpow3} it is not too difficult
to prove another restriction
on the coefficients $f^{(m)}_{i,j}$, namely
$f^{(m)}_{i>j,j}=0$.
\taglia{
 Indeed, from \eqref{ffpow3},  $r\geq i-1$, thus if
$i>j$, $r> (i+j)/2 -1$, thus $ 2 r > i+j-2$, thus $ r > i+j-2-r -p$
(since $p\geq0$). Therefore in \eqref{ffpow3} one has contributions
only from coefficients $f^{(m+1)}_{i',j'}$ with $i'>j'$.
Now, the total order associated to a coefficient
$f^{(m)}_{i,j}$ is $i+j$. From \eqref{ffpow3} it is evident that
coefficients $f^{(m)}_{i,j}$ with $o(m)=i+j$ depend on coefficients
$f^{(m+1)}_{i',j'}$ with $o(m+1)\leq o(m)-2$, and, iterating, the 
starting coefficients, i. e. the ones which does not come from the
integral, are $f^{(n)}_{0,0}=1$ with $n\leq m+o(m)/2$. These initial
coefficients are trivially vanishing when $i'>j'$, thus this rule
remains also to the lowest orders.

To evaluate $\abfr_{q}$ one needs to know the $\afr_i = f^{(0)}_{0,j}$ up to 
$j=q$. This depends on the $f^{(m)}$ with $o(m)\leq j-2 m \leq q-2 m
\leq q-2$. For a certain $o(m)$, the  maximum $p$ needed is $p\leq
o(m)-2$, i. e. $\abfr_{p'}$ with $p'=p+1\leq o(m)-2+1 \leq q-3 $.
}

We have evaluated the coefficients $f_j \equiv f^{(0)}_{0,j}$ up to $j = 42$, 
and, consequently, all other coefficients $f^{(m)}_{i,j}$, up to 
a corresponding high order,
by means of an algebraic manipulation program.
  If one tries to naively sum the series, one comes out
with apparent divergences already at $l$ of the order of one,
 which is probably 
 the convergence
radius. The Pad\'e method is most appropriate for
extrapolations beyond the convergence radius\cite{domb}. 
Indeed, we find that the possible poles are never on the
positive real
axis, neither close to it, which is the only region where we need
  $\FF_0(l)$ to be well defined. 
The simplest Pad\'e interpolation consists in
 equating the coefficients of the
series to the ones coming from 
 a rational function $P_n(l)/Q_k(l)$, where $P_n(l)$ is a
polynomial of order $n$, and $Q_k(l)$ one of order $k$. In
Fig.~\ref{padelog} we have plotted 
$\log P_n(l)/Q_k(l)$, as obtained by this result, with different 
$n,k$ close to  $20$. In all three cases, the logarithm seems to eventually 
acquire a constant slope, suggesting an exponential behavior for $\FF_0$. 
However, above $l\sim5$ the three different interpolations give different
results, which signals a failure of the Pad\'e procedure  for $l\cmag 5$.
A better approximation is achieved by 
making a rational interpolation to
the {\it logarithmic derivative}, i. e. to set the {\it Ansatz}
$\quad d \log[\FF_0(l)] /  \ d l =P_n(l)/Q_k(l)$.
As one can see from Fig.~\ref{padedlog},
 this logarithmic Pad\'e 
 interpolation now works good
 up to a larger $l \sim 20$ and it 
clearly shows that 
$d\ \log[\FF_0(l)] / \ d l \to 1$ for large $l$ (within about $10^{-4}$
of accuracy), i. e. that 
$\FF_0(l) \propto e^l$.

\levfig{
\begin{figure}
   \centerline{\psfig{file=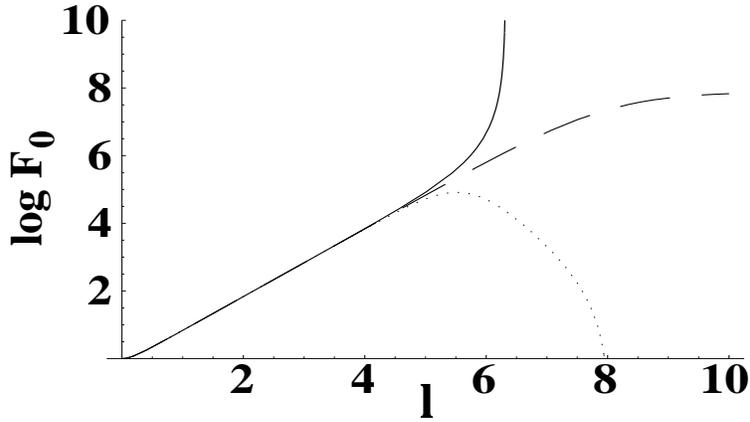,width=\figwidth}}
\caption{
\label{padelog}
$\log[\FF_0(l)]$ vs $l$ for $S=2$, obtained with a rational Pad\'e 
interpolation to the expansion of Sec.~\ref{s:solu} of the form
$ \FF_0(l)= P_n(l)/Q_k(l)$, with $n=k=20$ (solid line), $n=21,k=20$
(dashed), and $n=20,k=21$ (dotted).
}
 \end{figure}
}

\levfig{
\begin{figure}
   \centerline{\psfig{file=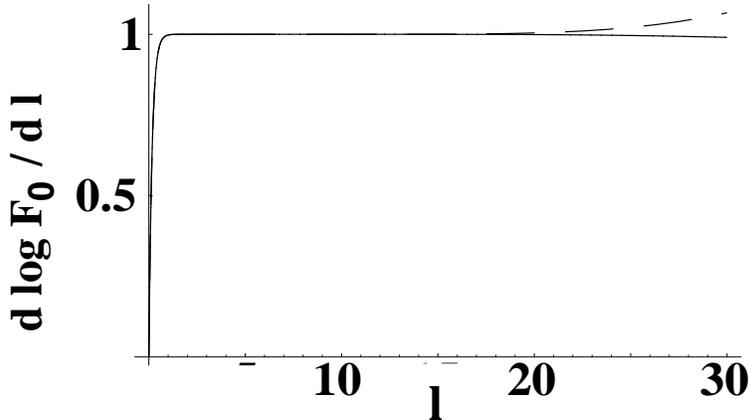,width=\figwidth}}
\caption{
\label{padedlog}
$d \ \log[F_0(l)]/d \ l$ vs $l$
with a logarithmic  Pad\'e approximation
$d\ \log[\FF_0(l)]/ \ d l =P_n(l)/Q_k(l)$, 
with $n=k=20$ (solid line), $n=21,k=20$
(dashed), and $n=20,k=21$ (dotted, covered by the solid line).
}
 \end{figure}
}

We also want
to study the behavior of the interaction vertices, given by
the 
RRC
\eqref{gc}. 
To this end, we have evaluated
their asymptotic behavior, when all internal variables $|\xt_k|$ are
of the same order of magnitude (see the discussion in Ref.~\onlinecite{rrc}).
This is obtained by setting all $l_k=l$ in \eqref{ff0}, or,
equivalently, 
$l\to m \ l$, and $l_m= l$ in \eqref{ff}.
In Fig.~\ref{pade35}, we have evaluated the logarithmic derivative of
$F_{m}(m \ l,l)$. The figure clearly shows that 
$F_{m}(m \ l,l)\propto e^{(m+1)\ l}$, i. e. the associated RRC, $\G_c$ of
\eqref{gc}, gets a correction proportional to $x^{(m+1) \al}$, where 
$x$ is the common value of all $|\xt_k|$.
Since the {\it bare} $m+1$-particle 
cumulant $\G^0_c$ scales like $x^{-(m+1)(1+\al)}$
(cf. Sec.~\ref{s:corr}), the anomalous exponent is again exactly canceled
by the renormalization. This is important, since one needs $\al$ to
scale to zero not only in the self--energy, but 
also in the interaction vertices, in order for the low--energy fixed 
point to be asymptotically free.

\levfig{
\begin{figure}
   \centerline{\psfig{file=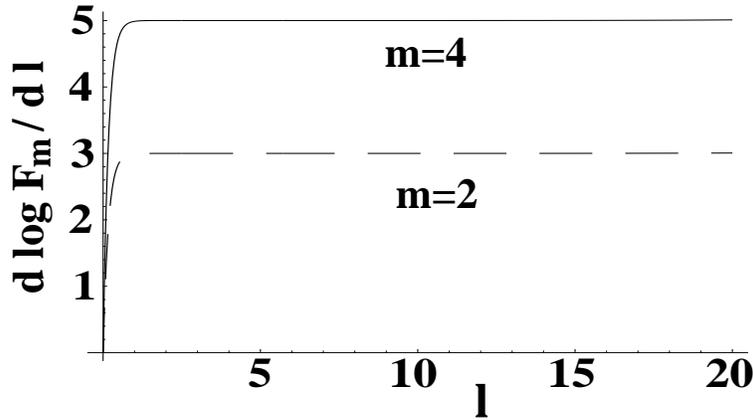,width=\figwidth}}
\caption{
\label{pade35}
$d \ \log[F_{m}(m \ l,l)]/d \ l$ vs $l$
for $m=4$ (solid line) and $m=2$ (dashed), 
with a logarithmic  Pad\'e approximation
$d\ \log[\FF_0(l)]/ \ d l =P_n(l)/Q_k(l)$ with $n=k=19$.
}
 \end{figure}
}

\taglia{
To conclude, we check that
the asymptotic 
solutions 
\beq
\label{asym}
\FF_{m}(l,l') \propto e^{  l   } \ e^{l'}
\eeq
 are indeed
consistent with the equation \eqref{ff}.
First, consider $\FF_0(l) = e^{l+a(l)}$, where, by hypothesis
 $a'(l) \to 0$ for large $l$.
Therefore, 
\beq
\label{barf0}
\bar \FF_0(l') + \bar FF_0'(l') = - a'(l') \ e^{-l'-a(l')} \;.
\eeq
Inserted in \eqref{ff}, the $e^{-l'}$ of \eqref{barf0}
 cancels the $e^{l'}$ of \eqref{asym}, thus there is no longer an
 $e^{l'}$ dependence in the integrand of \eqref{ff}. Thus, 
integrating over $l'$ only gives factors which increase less than
$e^{l_m}$. The only exponential contribution is thus  
$e^(l+l_m)$, which confirms the initial hypothesis.
}

\bibliography{footnoteslong,articles,preprints,mypublications,books}  

\begin{thebibliography}{10}
\bibitem{units}{
We collect here our conventions: (i) Units are chosen such that the equal
  charge and spin velocities $v_F$ and the Fermi energy $E_F$ are both equal to
  $1$. (ii) The momenta of the right- and left-moving fermions in the isolated
  LL are measured from the respective Fermi momenta, i. e. we set $k_F=0$.
  (iii) $\ww$ is the characteristic on-chain energy scale, i. e. the largest
  scale among the external frequency $\om$, the momentum $\kx$, and the
  temperature $T$. (iv) Coordinates (position $x$ or $y$, momentum $k$, etc.)
  in the direction parallel to the LL carry the subscript $\para$, while the
  ones in the ($D-1$-dimensional) plane perpendicular to it are labeled by
  $\perp$. (v) A boldface variable is a shorthand notation for (a) the variable
  in the $\para$ direction, (b) the corresponding imaginary-time or frequency
  variable, (c) the index $r$ indicating right- ($r=+1$) or left- ($r=-1$)
  moving electrons, and (d) the spin projection $\g$. Integrals over these
  coordinates implicitly include a sum over $r$ and $\g$. More specifically,
  $\xt$ stands for $(\xx,\tau,r,\g)$, $\yt$ for $(y_{\para},\tau,r,\g)$ (real
  space), and ${\kom}\equiv(\kx, -\om,r,\g)$ (momentum space). Notice that in
  momentum space, contrary to Ref.~\onlinecite{arri.99.c}, we have changed the
  sign of $\om$, so that $\kom \cdot \xt = \kx \xx - \om \tau$ ($r$ and $\g$
  don't enter the scalar product). In general, we use the same symbol for a
  function in real space and for its Fourier transform, if there is no
  ambiguity. (vi) The index $S$ indicates whether the system is spinless
  ($S=1$) or with spin ($S=2$). (vii) $\tp(\ky)\equiv \tp\cp$ is the kinetic
  energy in the $\perp$ direction, i. e. the Fourier transform of
  $\tp(\xy-\xy')$. For the simple case of a nearest--neighbor hopping
  $\tp/\resc$ considered here, the geometric factor $\cp$ is equal to $2
  \sum_{i=1}^{D-1} \cos(\ky_i)/\resc$. (viii) We rescale \tp by a factor
  $1/\resc$, with $D'=D-1$, in order to obtain a nontrivial $D\to \infty$
  limit, as pointed out in Ref.~\onlinecite{dinf}.}
\bibitem{compl}{
The region of LL behavior may extend down to quite low temperatures. For
  example, for typical parameters of the Bechgaard salts (see Ref. \cite{bech})
  like $\tp\sim 100 \ K$, $t_{\para}\sim 1000 \ K$, and, as it is believed
  $\al>1/2$, this region extends down to less than $10 \ K$. In the case of the
  Bechgaard salts, this is unfortunately already below the ordering transition.
  Moreover, for the above parameters one obtains, according to our result, a
  finite but considerably reduced quasiparticle weight $Z\sim 1/10$, which may
  be difficult to observe experimentally.}
\bibitem{local}{
We call $\perp$-local a quantity which is local in the $\xx$ coordinates or,
  equivalently, independent of the $\ky$ coordinates.}
\bibitem{great}{
A leading logarithmic contribution only comes from variables of different
  orders of magnitudes. In this sense, the inequalities ``$>$'' can be as well
  understood as ``$\gg$'', see Refs.~\onlinecite{russian,french}.}
\bibitem{rrc}{
The reason for the attribute ``restricted'' is that in \eqref{gc} we only
  consider contributions to the renormalized cumulant coming from the region
  $0\Reg {m+1}$. While, this eventually gives the leading contribution to the
  {\it single-particle} cumulant ($m=0$), it can be shown that this is not the
  case for the {\it many-particle} cumulants, unless all the $|\xt_k|$ are of
  the same order of magnitude.}
\bibitem{tdnolog}{
The reason is that in this region one can take the lowest order in the \tp
  expansion for $\td$, which gives \taglia{, after integration over $\cp$,
  $\td(\kom,\xy=0) \propto \tp^2 \ \Gamma(\kom)$, so that}
  $\td(\xt,\xy=0)\propto \tp^2 \ \Gamma(\xt) \propto \tp^2 \ |\xt|^{-1-\al}$.
  Inserting this result in \eqref{ffm}, one obtains for the integral restricted
  to this region $\int_1^{\teff^{-1}} |\xt|^{- 2 \al} \ d^2 \xt = O(1)$, i. e.
  no $\log$ contribution.}
\bibitem{ancont}{
To evaluate the spectral function one has to carry out an analytic continuation
  of the inverse self-energy to real frequencies. This is done by replacing
  $\iom\to z+i 0^+$ in \eqref{gamk1}. In the renormalization function $\FF_0$,
  this amounts to replace $|\kom| = \sqrt{w^2+\kx^2} \to \sqrt{k^2- (z+i
  0^+)^2}$.}
\end{thebibliography}
\bibliographystyle{myprsty} 


\end{document}